\documentstyle[sprocl,axodraw]{article}
\input epsf

\newcommand{\pr}{Phys. Rev. \underline}
\newcommand{\prl}{Phys. Rev. Lett. \underline}
\newcommand{\pl}{Phys. Lett. \underline}
\newcommand{\np}{Nucl. Phys. \underline}
\newcommand{\prep}{Phys.Reports \underline}

\newcommand{\MSbar}{\overline{\mbox{\small MS}}}
\newcommand{\mbar}{\overline{m}} 
 
\newcommand{\as}{\alpha_s}

\newcommand{\lqcd}{\Lambda_{{\rm QCD}}}
\newcommand{\beq}{\begin{equation}}
\newcommand{\eeq}{\end{equation}}
\newcommand{\be}{\begin{equation}}
\newcommand{\ee}{\end{equation}}
\newcommand{\beqn}{\begin{eqnarray}}
\newcommand{\eea}{\end{eqnarray}}
\newcommand{\bea}{\begin{eqnarray}}
\newcommand{\eeqn}{\end{eqnarray}}

\newcommand{\cale}{{\cal E}}

\newcommand{\oqcd}{O^{{\rm QCD}}}

\newcommand{\otwohqet}{O_2^{{\rm HQET}}}
\newcommand{\mbarb}{\overline m_b}
\newcommand{\msbar}{$\overline {{\rm MS}}$}

\newcommand{\plus}{\makebox[15pt][c]{$+$}}
\newcommand{\minus}{\makebox[15pt][c]{$-$}}
\newcommand{\err}[2]{\raisebox{0.08em}{\scriptsize
{$\;\begin{array}{@{}l@{}}
                          \plus\makebox[0.95em][r]{#1}
\\[-0.12em]
                          \minus\makebox[0.95em][r]{#2}
                        \end{array}$}}}
\def\slash#1{\mbox{$\not \!\! #1$}}

\def\spose#1{\hbox to 0pt{#1\hss}}
\def\ltapprox{\mathrel{\spose{\lower 3pt\hbox{$\mathchar"218$}}
 \raise 2.0pt\hbox{$\mathchar"13C$}}}
\def\gtapprox{\mathrel{\spose{\lower 3pt\hbox{$\mathchar"218$}}
 \raise 2.0pt\hbox{$\mathchar"13E$}}}
\def\inapprox{\mathrel{\spose{\lower 3pt\hbox{$\mathchar"218$}}
 \raise 2.0pt\hbox{$\mathchar"232$}}}
\newcommand{\dslash}{\slash{\partial}}

\newcommand{\nn}{\nonumber}
\newcommand{\<}{\langle}
\renewcommand{\>}{\rangle}

\begin{document}
\begin{flushright}
CERN-TH/96-130\\
SHEP 96-13\\
\end{flushright}

\title{LATTICE SIMULATIONS AND EFFECTIVE THEORIES}

\author{C.T.SACHRAJDA}

\address{Theory Division, CERN, CH-1211 Geneva 23, Switzerland, and \\
  The Department of Physics, University of Southampton, Southampton,
  SO17 1BJ, UK}

\maketitle\abstracts {I present a brief introduction to the lattice
  formulation of quantum field theory, and discuss the use of lattice
  simulations for studies in particle physics phenomenology. The
  computation of $f_B$, the decay constant of the $B$-meson, is used
  as a case study. I also explain the appearance and cancellation of
  ``renormalons'' in the evaluation of power corrections (higher-twist
  corrections) in hard scattering and decay processes.}

\paragraph{Introduction}\ \ 

The lattice formulation of quantum field theories, together with large
scale numerical simulations, is becoming a very powerful
non-perturbative tool in many areas of particle physics, and in the
evaluation of long-distance strong-interaction effects in physical
processes, in particular.  In these lectures I will present a brief
introduction to lattice computations in QCD (lecture 1) and will then
discuss, as a case study, the applications to the evaluation of the
decay constant of the $B$-meson, and other physical quantities, in the
Heavy Quark Effective Theory (HQET).  Finally in the third lecture I
will review the question of ``renormalons'', which arises when one
evaluates power corrections (i.e. higher-twist corrections) to hard
scattering and decay processes.

The principal objective of these lectures is to explain the
theoretical framework needed for lattice simulations in particle
physics, and to give some idea of what is possible now and what the
outstanding problems are. In the first two lectures I will illustrate
the discussion with numerical results obtained by the UKQCD
collaboration, of which I am a member. It is not my intention, nor
would it be appropriate at this school, to present a critical review
of the many interesting results which have appeared in the last few
years from many different groups. The proceedings of the annual
symposia on Lattice Field Theory contain detailed reviews and status
reports on all aspects of the subject, and provide very helpful
starting points for literature searches. A selection of review talks,
which are suitable as further reading to the material of these lectures,
can be found in refs.~\cite{latt93}$^-$\cite{melbourne}.

\section{Lecture 1 -- Lattice Computations of Masses and Matrix Elements}
\label{sec:lattice}

In this first lecture I will briefly review the ingredients used in
evaluating hadronic masses and matrix elements in QCD. Simulations
using the Heavy Quark Effective Theory will be described in the
remaining two lectures. 

The quantity which is evaluated directly in lattice computations is
the vacuum expectation value of a multi-local operator 
$O(x_1,x_2,\cdots ,x_n)$, composed of quark and gluon fields:
\be
\< 0|\, O(x_1,x_2,\cdots ,x_n)\,|0\> = 
\frac{1}{Z}\int [dA_\mu][d\psi][d\overline{\psi}]\,
e^{iS} O(x_1,x_2,\cdots ,x_n) \ ,
\label{eq:vev}\ee
where $Z$ is the partition function 
\be
Z=\int [dA_\mu][d\psi][d\overline{\psi}]\, e^{iS}
\label{eq:Z}\ee
and $S$ is the QCD action. In these introductory paragraphs all the
formulae are written in Minkowski space, although, of course, the
lattice simulations use a Euclidean formulation of QCD. For the
computations described in these lectures, $n$ will be either equal to
2 or 3, and we now describe the significance of the two- and
three-point correlation functions respectively.

\paragraph{Two-Point Correlation Functions:} 
Consider the bilocal operator $O$:
\be
O(x_1,x_2) = J(x_1)\,J^{\dagger}(x_2)\ ,
\label{eq:o2def}\ee
where $J$ and $J^{\dagger}$ are interpolating operators which can
destroy or create the hadron $H$, whose mass will be denoted by $m_H$.
Moreover, we will assume that $H$ is the lightest hadron which can be
created by $J^\dagger$. We now define the two-point correlation
function:
\be
C_2(t) = \int d^3x\, e^{i\vec p\cdot\vec x}
\< 0\,|\, J(\vec x,t)\, J^\dagger(\vec 0,0)\,|\,0\>\ ,
\label{eq:c2def}\ee
where we take $t>0$. Inserting a complete set of states, $\{n\}$,
between the two operators we obtain 
\beqn
C_2(t) & = & \sum_n \int d^3x\, e^{i\vec p\cdot\vec x}
\< 0\,|\, J(\vec x,t)\,|\,n\>\<\,n| J^\dagger(\vec 0,0)\,|\,0\>
\\ 
& = & \frac{1}{2E} |\<0\,|\,J(\vec 0,0)\,|H(p)\>|^2\,e^{-iEt} + \cdots\ ,
\label{eq:c2states}\eeqn
where the ellipses represent contributions from states which are
heavier than $H$, and we have used translational invariance to
displace the argument of $J$ from $(\vec x,t)$ to the origin. The
four-momentum $p$ has components $(\vec p, E)$, 
where $E^2=\vec p^{\, 2} + m_H^2$. In Euclidean
space the exponential factor in eq.(\ref{eq:c2states}) is $e^{-Et}$,
and the contributions of the heavier states have similar 
exponential factors. Thus for sufficiently large
times $t$, the contribution from the heavier states is suppressed, and
the ground state is isolated. By fitting the time behaviour of the 
computed correlation function, one obtains $E$, and hence the mass of the
hadron $H$, and also the matrix element $|\<0\,|\,J(\vec 0,0)\,|H(p)\>|$.
For example if $H$ is the pion, and $J$ the axial current, then from this
matrix element we can obtain the value of the pion's decay constant:
\be
|\<\, 0 \,|\,A_\mu(0)\,|\pi(p)\,\>| = f_\pi\,p_\mu  \ .
\label{eq:fpidef}\ee

\paragraph{Three-Point Correlation Functions:} 
It will also be useful for us to consider three-point correlation functions:
\be
C_3(t_x, t_y) = \int d^3x\, d^3y\, e^{i\vec p\cdot\vec x}
e^{i\vec q\cdot\vec y}
\< 0\,|\,J_2(\vec x, t_x)\, O(\vec y,t_y)\, J^\dagger_1(\vec 0, 0)\,
|\, 0\>\ ,
\label{eq:c3def}\ee
where, $J_1$ and $J_2$ are the interpolating operators for hadrons
$H_1$ and $H_2$ respectively, $O$ is a local operator, and we have assumed
that $t_x>t_y>0$. Inserting complete sets of states between
the operators in eq.(\ref{eq:c3def}) we obtain
\beqn
\lefteqn{C_3(t_x, t_y) = 
\frac{e^{-iE_1t_y}}{2 E_1}\ \frac{e^{-iE_2(t_x - t_y)}}{2 E_2}\,
\< 0\,|\,J_2(\vec 0, 0)\,|\,H_2(\vec p, E_2)\>\times}\nn\\ 
& & \< H_2(\vec p, E_2)\,|\,O(\vec 0,0)\,|\,H_1(\vec p + \vec q, E_1)\>\,
\< H_1(\vec p + \vec q, E_1)\,|\,J^\dagger_1(\vec 0,0)\,|\,0\>+\cdots\ ,
\label{eq:c3states}\eeqn
where $E_1=\sqrt{m_1^2 + (\vec p+\vec q)^2}$, $E_2=\sqrt{m_2^2+\vec
  p^2}$ and the ellipses represent the contributions from heavier
states. In Euclidean space the exponential factors become
$\exp(-E_1t_y)$ and $\exp(-E_2(t_x-t_y))$, so that for large time
separations, $t_y$ and $t_x-t_y$, the contributions from the lightest
states dominate.  All the elements on the right-hand side of
eq.(\ref{eq:c3states}) can be determined from two-point correlation
functions, with the exception of the matrix element $\<H_2|O|H_1\>$.
Thus by computing two- and three-point correlation functions the
matrix element $\<H_2|O|H_1\>$ can be determined.

The computation of three-point correlation functions is useful in
studying weak decays of hadrons (e.g. if $H_1$ is a $D$-meson, $H_2$ a
$K$ meson and $O$ the vector current $\overline{c}\gamma^\mu s$, then
from this correlation function we obtain the form factors relevant for
semileptonic $D\to K$ decays) and in determining properties of
hadronic structure (e.g. if $H_1$ and $H_2$ both represent the proton
and $O$ is the electromagnetic current, then from $C_3$ we obtain the
electromagnetic form-factors of the proton).

The computation of correlation functions will be outlined in 
subsec.~\ref{subsec:cfeval}, but I start with a brief description of the 
formulation of QCD on a discrete space-time lattice.

\subsection{Elements of Lattice QCD}
\label{subsec:latticeqcd}
In order to perform lattice computations it is necessary to formulate
QCD in discrete space-time and in Euclidean space~\footnote{For
  excellent pedagological introductions to the lattice formulation of
  QCD see refs.\cite{mm,creutz}.}. The quark fields $\psi(x)$ are
field variables defined on the sites, \{$x$\}, of the lattice. The gluon
fields are conveniently introduced in terms of ``link'' variables
$U_\mu(x)$, defined on the link between the point $x$ and $x + a
\hat\mu$, where $\hat\mu$ is the unit lattice vector in the $\mu$
direction. The gluon field $A_\mu$, is then given by
\be 
U_\mu(x) = e^{iag_0A_\mu(x+a\,\hat\mu/2)}\ ,
\label{eq:adef}\ee
where $a$ is the lattice spacing and $g_0(a)$ is the bare coupling constant
defined with the lattice action being used. Under a local gauge transformation
$g(x)$, the fields transform as:
\be
\psi(x)\to g(x)\psi(x)\hspace{0.4in}{\mathrm and}\hspace{0.4in}
U_\mu(x)\to g(x)U_\mu(x)g^\dagger(x+a\hat\mu)\ .
\label{eq:gaugetr}\ee
$U^\dagger_\mu(x-a\hat\mu)$ can be thought of as the link variable from
$x$ to $x-a\hat\mu$. From eq.(\ref{eq:gaugetr}) it can be seen that
any closed loop of link variables will be gauge invariant, and this is
exploited in constructing the action and the operators which represent 
observables.
For example, the Wilson action for the gauge fields is defined in terms of
the ``plaquette'' variable, which is the product of link variables
around an elementary square of the lattice:
\be
{\cal P}_{\mu,\nu}(x) \equiv {\mathrm Tr}\,[U_\mu(x)U_\nu(x+a\hat\mu)
U^\dagger_\mu(x+a\hat\nu)U^\dagger_\nu(x)]\ .
\label{eq:plaquettedef}\ee
The Wilson action is defined by
\be
S_{{\mathrm gluon}}\equiv \beta\sum_{x,\mu,\nu}\left[ 1 - \frac{1}{3}
{\mathrm Re}\, {\cal P}_{\mu,\nu}(x)\right]\ ,
\label{eq:gluonaction}\ee
where $\beta=6/g_0^2$. Expanding the right hand side of
eq.(\ref{eq:gluonaction}) in powers of the lattice spacing, one can
verify that, as required, it gives $\int d^4x \frac{1}{4} F_{\mu\nu}^2$,
up to ``irrelevant'' terms of $O(a^2)$.

When trying to discretise the fermion action, one encounters the
famous problem of ``Fermion Doubling''. For example consider the free
quark propagator, corresponding to the Dirac action $\overline{\psi}(
\dslash + m)\psi$. Defing the derivative of a function as the difference
of its values at neighbouring sites, one
obtains the following lattice propagator in momentum space
\be
S(q)_{{\mathrm free}} = m+\frac{i}{a}\sum_\mu
\gamma_\mu \sin (a q_\mu)\ .
\label{eq:sfree}\ee
Consider now $q$ satisfying the continuum mass-shell condition
$q^2+m^2=0$. The free-propagator then has a pole at this choice of $q$
as expected (modified by terms which vanish as $a\to 0$). Now there
are also poles at the values of momenta with any component $q_\mu$
replaced by $\pi/a- q_\mu$, leading to sixteen states instead of one.
Such doubling of the fermion degrees of freedom is a general feature
of lattice theories \cite{nn}. Wilson's solution to this problem was
to add an ``irrelevant'' operator to the action, in such a way that
the unphysical doublers have infinite masses as $a\to 0$. For the free
theory the Wilson term takes the form: 
\beqn -\frac{ra}{2}
\overline{\psi}\partial^2\psi & = & -r a^4\sum_x\Big\{
\frac{1}{2a}\sum_\mu[\overline{\psi}(x)\psi(x+a\hat\mu)
\nn\\
& + &
\overline{\psi}(x)\psi(x-a\hat\mu)-2\overline{\psi}(x)\psi(x)]\Big\}\ ,
\label{eq:wilsonterm}\eeqn
where $r$ is an arbitrary parameter. Physical quantities must, of
course, be independent of $r$. With the Wilson term, the free propagator
becomes
\be
S(q)_{{\mathrm free}} = m+\frac{i}{a}\sum_\mu
\gamma_\mu \sin (a q_\mu) + \frac{r}{a}\sum_\mu(1-\cos(aq_\mu))\ .
\label{eq:sfreew}\ee
from which the shift of the mass of the doublers can be deduced. The Wilson
term breaks chiral symmetry, even for $m=0$, but the symmetry is restored at
a special value of the bare mass~\cite{bochicchio} (see below). Including
now interactions between quarks and gluons, the Wilson action becomes:
\beqn
S_W  & = & 
S_{{\mathrm gluon}} + a^4\sum_x\Big\{ -\frac{\kappa}{a} 
\sum_\mu\left[\overline{\psi}(x)(r-\gamma_\mu)U_\mu(x)\psi(x+a\hat\mu)
+\right. \nn\\ 
& & 
\left.
\overline{\psi}(x+a\hat\mu)(r+\gamma_\mu)U^\dagger_\mu(x)\psi(x)\right]
+\overline{\psi}(x)\psi(x)\Big\}\ ,
\label{eq:wilsonaction}\eeqn
where the parameter $\kappa=1/(2m + 8r)$ contains the 
dependence on the mass of the quark. The fields have been rescaled by
a factor of $2\kappa$, and of course, a new term must be included for
every quark flavour (each with a different $\kappa$ parameter). 

I end this subsection with some brief comments:
\begin{enumerate}
\vspace{-8pt}
\item[i)] The only parameters of the theory are those of QCD itself,
  i.e. the coupling constant, and the quark masses (parametrised in
  terms of $\kappa$).
\item[ii)] As mentioned above, physicial quantities must be
  independent of $r$, and we set $r=1$ in most of the following
  discussion.
\item[iii)] There is a value of $\kappa$ ($=\kappa_c$) for which
  chiral symmetry is restored, ($\kappa_c = 1/8 +
  O(\as)$)~\cite{bochicchio}.
\item[iv)] Expanding the right hand side of eq.(\ref{eq:wilsonaction}) in
powers of the lattice spacing we find
\beq
S_W = S_{QCD}^{{\mathrm cont}} + O(a)\ ,
\label{eq:oa}\ee
where the superscript ``cont'' implies that this is the continuum QCD
action.  Thus, for a finite lattice spacing $a$, we expect there to be
``discretisation'' errors of $O(a)$ (i.e. of $O(pa), O(ma)$ or of 
$O(\lqcd a)$, where $p$ is the momentum of one of the hadrons). Much work
is currently being done in trying to reduce these lattice artefacts, by
using ``improved'' lattice actions and operators~\cite{symanzik}.
\end{enumerate}

\subsection{Evaluation of Correlation Functions}
\label{subsec:cfeval}

In this subsection I will briefly describe the steps used in the 
evaluation of correlation functions, without explaining the algorithms which
are used in the computations. 

\paragraph{Pure Gauge Theories:} In the pure gauge theory, and in Euclidean
space the functional integral in eq.(\ref{eq:vev}) becomes:
\be
\< 0|\, O(x_1,x_2,\cdots ,x_n)\,|0\> = 
\frac{1}{Z}\int [dU]
e^{-S(U)} O(x_1,x_2,\cdots ,x_n) \ ,
\label{eq:vevpg}\ee
where $Z$ is given by
\be
Z=\int [dU] e^{-S(U)}
\label{eq:Zpg}\ee
and $S$ is the gluon action (\ref{eq:gluonaction}). Examples of
interesting studies in pure gauge theories include the evaluation of
the spectrum of glueball states and the detemrination of the potential
between a static quark and an anti-quark (for a recent review, and
references to the original literature, see ref.~\cite{glueballs}).

On a finite lattice with $V$ points, the right
hand side of eq.(\ref{eq:vevpg}) is a well-defined multi-dimensional
integral (specifically, for an $SU(n)$ gauge theory it is a $4 V
(n^2-1)$ dimensional integral). Monte Carlo algorithms are an efficient
way to evaluate these multi-dimensional integrals. With these algorithms
gluon configurations are generated on all the links of the lattice, with
a probability density $1/Z \exp(-S(U))$. Thus 
\be
\< O \> = \frac{1}{N_c}\sum_{c=1}^{N_c}\,O(U_c) \,
\label{eq:average}\ee
where $O(U_c)$ is the value of $O$ for the field configuration $\{U_c\}$, 
and the sum is over $N_c$ statistically independent gluon
configurations.  In general we do not very know much about the statistical
distribution of the results, but expect that the uncertainty will typically 
decrease like $1/\sqrt{N_c}$. The ``statistical'' error for a quantity
in a given computation is estimated from the variation of the result 
as gluon configurations are subtracted and added. 

\paragraph{Lattice QCD}
We now write the Wilson action (\ref{eq:wilsonaction}) in the
following shorthand notation:
\beq
S_W=\bar\psi\,\Delta(U)\psi + S_{\mathrm gluon}
\label{eq:swsh}\ee
wher $S_{\mathrm gluon}$ is the gluon action and $\Delta$ is a matrix in position, spin and 
colour spaces. The functional integral over the quark fields is 
straightforward to evaluate formally:
\beqn
\int [d\psi][d\bar\psi]\,e^{S_W} & = & {\mathrm det}[\Delta(U)] \,
e^{-S_{\mathrm gluon}}
\label{eq:ffi1}\\ 
\int [d\psi][d\bar\psi] \psi_i \bar\psi_j\,
e^{S_W} & = & \Delta^{-1}_{ij}(U)\, {\mathrm det}[\Delta(U)]\,
e^{-S_{\mathrm gluon}}
\label{eq:ffi2}\eeqn
etc. The labels $i$ and $j$ represent the lattice coordinates, as well
as spin and colour indices. $\Delta^{-1}(U)$ is the quark propagator
in the background field $\{U\}$. Thus for the integration over the
gluon fields, the action should now be modified to $S_{\mathrm
  gluon}-\ln {\mathrm det}[\Delta(U)]$.  The second term, which
represents the effects of closed quark loops, is non-local, and its
inclusion in the Monte-Carlo algorithms is very expensive in terms of
computing resources. It is the focus for much of the development work
in improved algorithms for lattice simulations. At present, most large
scale numerical simulations are performed in the ``quenched
approximation'', in which the determinant is not included, i.e.  the
effects of closed quark loops are neglected. Even where they are
explicitly included, it is generally the case that the sea quarks are
heavy (with masses about those of the strange quark), so that it is
likely that the effects of light quark loops are significantly
underestimated in these studies.

We do not really know how large the errors introduced by the quenched
approximation are, and expect that these will depend on the physical
quantities being computed, as well as on the quantities being used to
fix the lattice spacing (and quark masses). Over the last ten years
and more, a wide variety of quantities have been computed in quenched
simulations and compared to the experimental values. In almost all
cases there is good agreement between the computed and measured values
(of the order or better than 25\% or so), giving us confidence in the
predictions for unknown physical quantities, such as the decay
constant $f_B$ discussed in the second lecture.  The errors due to
quenching are likely to be larger when one compares physical
quantities which depend on different scales (e.g.  if one fixes the
lattice spacing in some process which occurs at a scale of $O(\lqcd)$
and uses it in the prediction for heavy quarkonia). Of course it is
only when we are able to include light quark loops in a reliable way
that we will be able to claim that lattice QCD is a truly quantitative
non-perturbative technique.

For the rest of these lectures I will discuss results obtained 
with the quenched approximation.

\subsection{The Pion Propagator}
\label{subsec:pionprop}

\begin{figure}[t]
\begin{center}
\begin{picture}(180,60)(-90,-45)
\Oval(0,0)(30,60)(0)
\Vertex(-60,0){3}\Vertex(60,0){3}
\ArrowLine(-0.5,30)(0.5,30)
\ArrowLine(0.5,-30)(-0.5,-30)
\Text(-80,0)[c]{$\gamma^0\gamma^5$}
\Text(80,0)[c]{$\gamma^0\gamma^5$}
\Text(-60,-40)[c]{0}
\Text(60,-40)[c]{$t$}
\LongArrow(120,-40)(150,-40)
\Text(135,-35)[b]{time}
\end{picture}
\caption{Diagramatic representation of the two-point correlation
  function of the pion, eq.(\protect\ref{eq:pionprop}). The lines
  represent quark propagators in the gluon background field.}
\label{fig:pionprop}
\end{center}\end{figure}
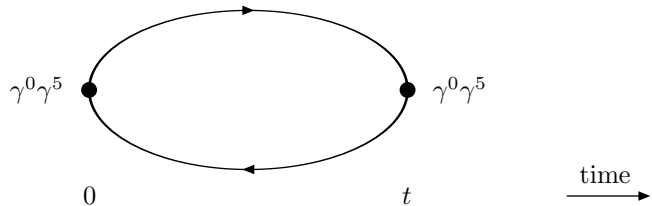

I conclude this lecture with a specific example, that of the
evaluation of the pion's decay constant $f_\pi$. The evaluation of other
matrix elements and hadronic masses follows similar steps. Consider now
the correlation function of two axial currents:
\beqn
C(t) &\equiv & \sum_{\vec x}\<\,0\,|\bar\psi(0)\gamma^0\gamma^5\psi(0)\ 
\bar\psi(x)\gamma^0\gamma^5\psi(x)\,|\,0\,\>\\ 
& = & - \frac{1}{Z}\,\int\, [dU]\,e^{-S_{\mathrm gluon}}\, {\mathrm Tr}
[S(0,x)\gamma^0\gamma^5 S(x,0)\gamma^0\gamma^5]\ ,
\label{eq:pionprop}\eeqn
where $x=(\vec x, t)$ and I have denoted the quark propagator (with
the dependence on the background field implicit) by $S(x,0)$. This
correlation function is represented by the diagram in
fig.~\ref{fig:pionprop}.

The symmetry property $S(0,x)=\gamma^5\,S^\dagger(x,0)\gamma^5$ is
particularly useful, as it implies that in order to evaluate the
correlation function $C(t)$, for each gluon configuration it is
sufficient to determine the set of quark propagators from an arbitrary
lattice point to the origin, $\{S(x,0)\}$. This is achieved by using
efficient algorithms for the inversion of sparse matrices. 
I now assume that these sets of
propagators have been generated for many gluon configurations and that
$C(t)$ has been evaluated.

\begin{figure}[t]
   \epsfxsize=8cm
   \centerline{\epsffile{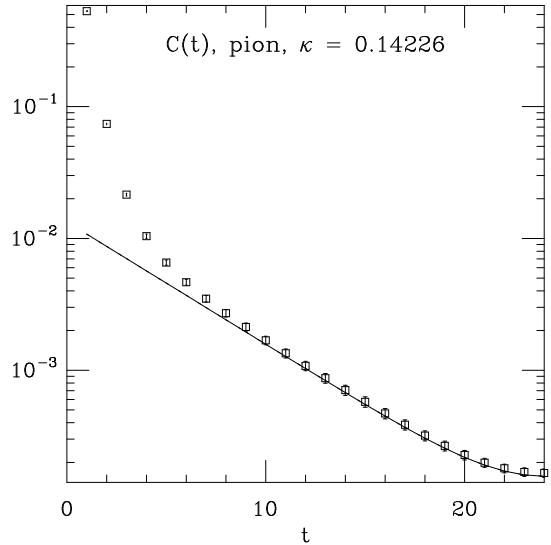}}
   \caption{\label{fig:ukqcdpi}
     Correlation function of the light pseudoscalar meson (``pion'')
     with quarks which are a little lighter than the strange quark
     (from ref.~\protect\cite{ukqcdstrange}).}
   \end{figure}

Fig.~\ref{fig:ukqcdpi} shows the pion's correlation function obtained
on a $24^3\times 48$~lattice at $\beta=6.2$ (for which the inverse
lattice spacing is about 2.9 GeV), using 60 gauge field configurations
(and the SW fermion action, which is an ``improvement'' on the Wilson
action as explained in section~\ref{subsec:propagating}).  The light
quarks are a little lighter than the physical mass of strange quark.
In the same figure is shown the fit to a single state. The fit allows
for the contribution of the backward propagating meson (which arises
because, with periodic boundary conditions, $t$ is both greater than
and smaller than 0, so that there are two contributions for any $t$),
so that the exponential is replaced by a hyperbolic cosine (which
makes a difference close to the centre of the lattice).

\begin{figure}[t]
  \epsfxsize=8cm \centerline{\epsffile{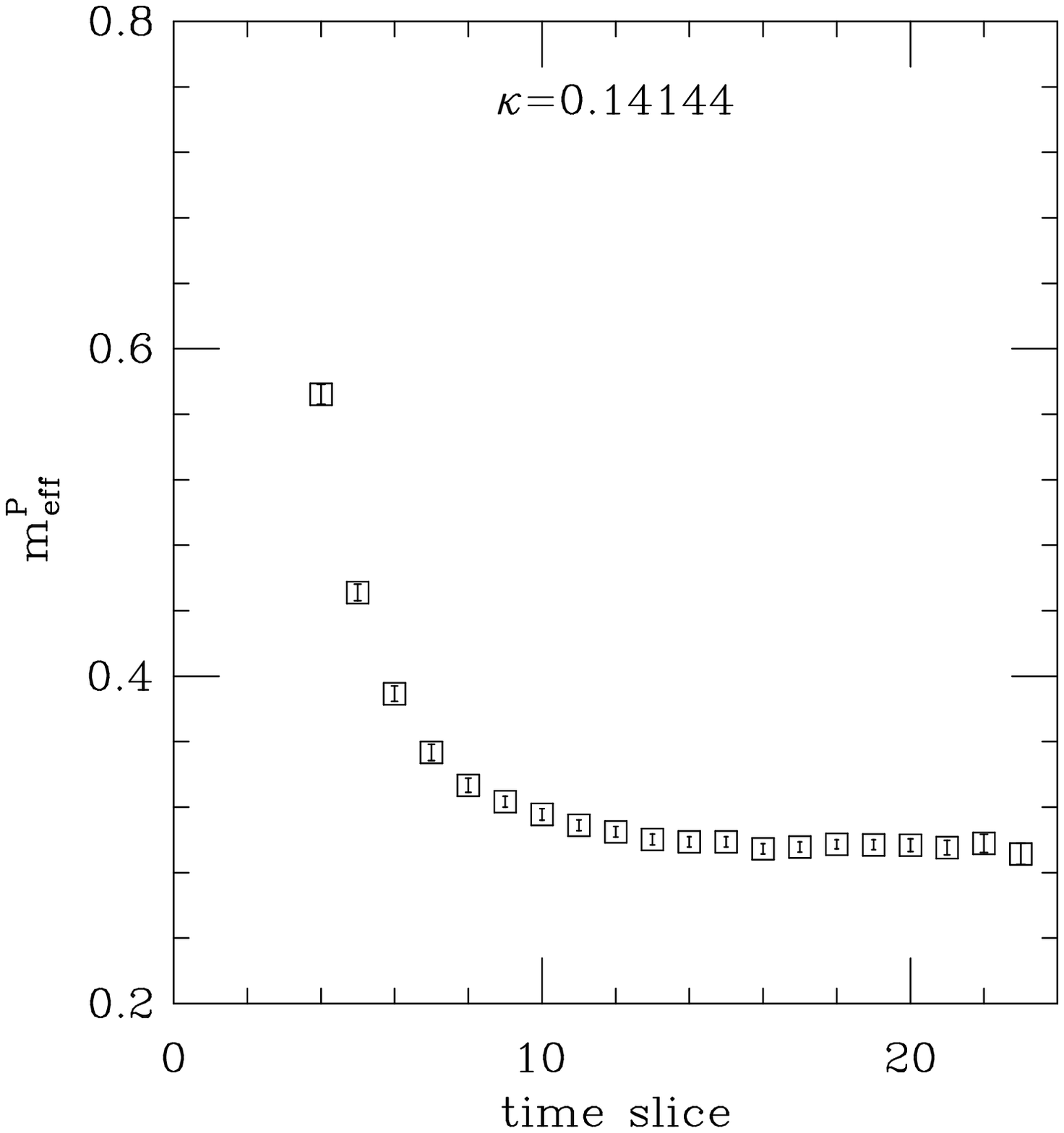}}
  \caption{\label{fig:effmass} The effective mass plot of the light 
      pseudoscalar meson (``pion'') with quarks which now are a little 
      heavier than the strange quark (from 
      ref.~\protect\cite{ukqcdstrange}). }
\end{figure}

Following the discussion above we need to establish whether we can
isolate the ground state, i.e. whether the correlation functon is well
represented by a single exponential over some interval at larger
times. To this end it is useful to define the effective mass by
\be
m_{{\mathrm eff}}(t)\equiv\ln\left[\frac{C(t)}{C(t+1)}\right]\ ,
\label{eq:meffdef}\ee
so that if the correlation function is well represented by a single
exponential, then $m_{{\mathrm eff}}$ will be almost independent of
the time $t$. The criteria for the construction of good lattice
interpolating operators, such that the overlap with the lightest state
is enhanced, and a plateau in the plot of $m_{{\rm eff}}(t)$ as a
function of $t$ is reached at relatively small times, is an area under
intensive investigation. In order to determine $f_\pi$ however, we
need to determine the matrix element of a local axial current. The
effective mass plot for the $\pi$-mesons, now for light quark masses
which are a little heavier than the strange quark, obtained by the
UKQCD collaboration with local interpolating operators are presented
in fig.~\ref{fig:effmass}.

\begin{figure}[t]
  \centerline{\epsfxsize=6cm\epsffile{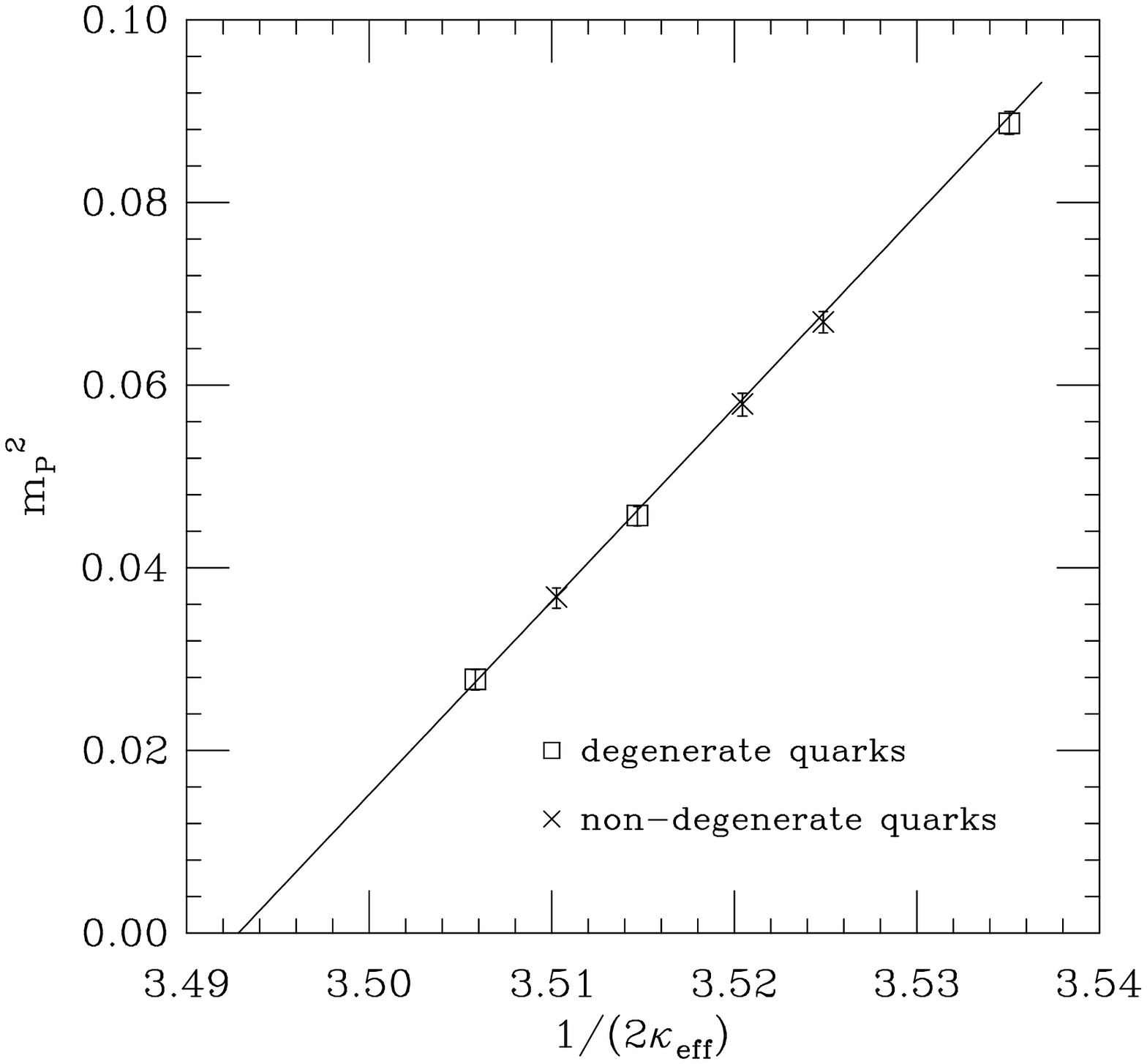}
    \vspace{10pt}\epsfxsize=6cm\epsffile{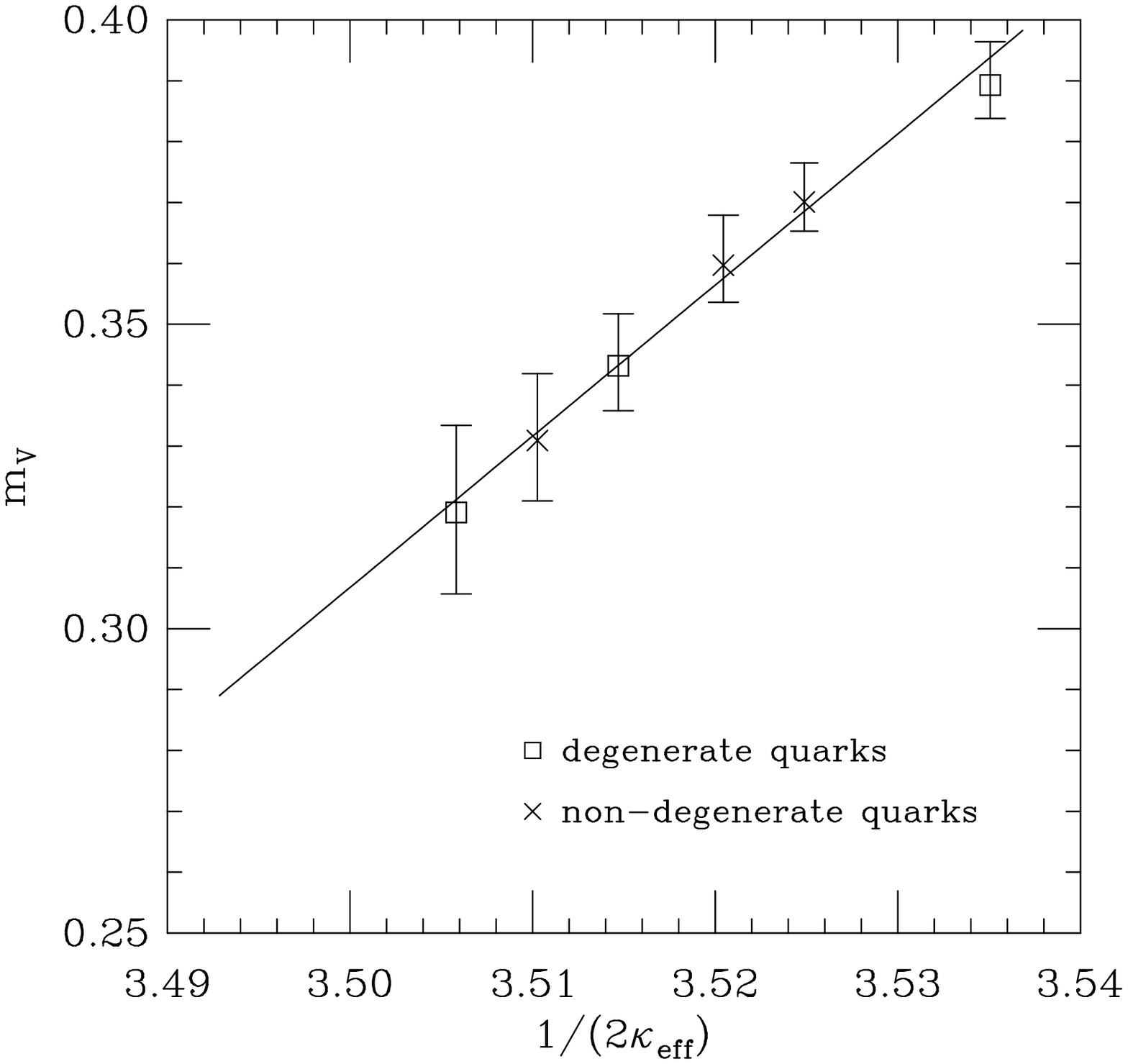}}
  \caption{\label{fig:chiral}
        The values of (a) the pion mass squared, and (b) the rho
        mass plotted as a linear function of the quark
        mass(from ref.~\protect\cite{ukqcdstrange}).}
\end{figure}

Because of finite volume effects it is not possible to perform the
computations with physical light-quark masses, rather one must
extrapolate the results obtained for masses around that of the strange
quark to the chiral ($m_q=0$) limit. This requires the knowledge of
the value of the $\kappa$-parameter at the chiral limit ($\kappa_c$).
To determine $\kappa_c$, we use PCAC and plot $m_\pi^2$ as a function
of $1/\kappa$ (and hence as a linear function of the quark mass), and
determine the value of $\kappa$ at which $m_\pi^2=0$ (or $m_\pi$ takes
its physical value). In fig.~\ref{fig:chiral}(a) we show such a plot,
again from the UKQCD collaboration's simulation at $\beta=6.2$. In
fact it can be shown using chiral perturbation theory that the PCAC
relation stating that $m_\pi^2$ is proportional to the quark mass is
violated in quenched simulations \cite{sharpe,bg}. However this effect
only occurs at very low quark masses which is difficult to simulate
without encountering finite volume effects. Recently Kim and Sinclair
have demonstrated this violation of the PCAC in a numerical simulation
on very large lattices \cite{ks}.

Having determined $\kappa_c$ we now extrapolate the results for
$f_\pi$ to the chiral limit. Generally, in the absence of any
theoretical guidance, one assumes a linear dependence on the mass of
the light quark in this extrapolation. Now $f_\pi$ is a dimensionful
quantity, which we determine in lattice units, i.e. we compute
$af_\pi$, where $a$ is the lattice spacing. Thus to make a prediction
for $f_\pi$ we need to know the value of $a$. In lattice computatons
this is done implicitly using dimensional transmutation. Instead of
fixing $a$, and then using some physical quantity to determine the
bare coupling constant $g_0(a)$ (which is a genuine parameter of QCD),
one fixes $g_0(a)$ (or equivalently $\beta$) and determines $a$. For
example one may use the mass of the $\rho$-meson to determine $a$ (an
extrapolation to the chiral limit of the UKQCD results, at $\beta =
6.2$, is presented in fig.~\ref{fig:chiral}(b)).  My summary of the
values of the lattice spacings for two commonly used values of $\beta$
(obtained by using light quark quantities to set the scale) is
\beqn
a^{-1}(\beta=6.0) & = & 2.0(2)\, {\mathrm GeV}\label{eq:a6pt0}\\
a^{-1}(\beta=6.2) & = & 2.9(3)\, {\mathrm GeV}\ .
\label{eq:a6pt2}\eeqn
The errors are largely due to the variation in the results obtained
using different physical quantities to set the scale, and hence are at
least partially due to quenching. An inverse lattice spacing of 2~GeV
corresponds to a spacing of 0.1~fm.

In this way one obtains the matrix element of the axial current (or
similarly other matrix elements) from lattice simulations. The matrix
element is of a bare operator which has been defined in QCD with a
lattice regularization, and a specific lattice action. The
ultraviolet properties of such an operator are of course different
from those defined in the continuum. However, since it is the
short-distance behaviour which is different, the relation between bare
lattice operators defined with an inverse lattice spacing
$a^{-1}\gg\lqcd$ and renormalized continuum operators defined at a scale
$\mu\gg\lqcd$, can be derived using perturbation theory~\footnote{ For
  some quantities this matching between the lattice and continuum
  operators represents the largest source of theoretical uncertainty.
  Non-perturbative methods to evaluate the matching factors are being
  developed~\cite{nonpertren1,nonpertren2}.}.

As an example of the results obtained for $f_\pi$, the UKQCD
collaboration, from the simulation which was used to illustrate the
calculational steps above, quote $f_\pi/m_\rho = 0.14(1)$, to be
compared to the experimental result of 0.17.

\section{Lecture 2 -- Simulations in The Heavy Quark Effective Theory}
\label{sec:hqet}

In this lecture I will apply the formalism reviewed above to lattice
simulations in the HQET, using the evaluation of the decay constant of
the $B$-meson ($f_B$) as a case study.  Matthias Neubert has, at this
school, already descibed the formalism of the Heavy Quark Effective
Theory (HQET) and explained its usefulness \cite{mnalmunecar}. In this
framework, predictions for physical quantities in heavy quark physics
are presented in terms of a series in inverse powers of the mass of
the heavy quark. The predictions are given in terms of matrix elements
of operators containing the field of the heavy quark, and lattice
simulations provide the opportunity for the evaluation of these matrix
elements.

Before starting the presentation however, I would like to discuss why
it useful to perform lattice computations in the HQET, rather than to
evaluate quantities such as $f_B$ directly in QCD. The reason is that
the sizes of the lattices and lattice spacings which we can be used
are limited by the available computing resources, and current quenched
simulations are performed with inverse lattice spacings typically in
the range of 2--3.5~GeV. Thus the lattice spacings are larger than the
Compton wavelength of the $b$-quark, and we cannot meaningfully study
its propagation directly in QCD. Since in lattice computations we have
the luxury of varying the masses of the quarks, one approach to the
study of $B$-physics on the lattice is to perform the computations
with several smaller values of the heavy quark mass (typically in the
range of that of the charm quark mass), and to extrapolate the results to
the $b$-quark\,\footnote{In the following I shall refer to such
  simulations as ones with ``propagating'' heavy quarks, in
  distinction to the ``static'' quarks of the HQET.}. Of course such
an extrapolation introduces uncertainties, and much effort is being
spent on reducing the lattice artefacts (errors due to the finite size
of the lattice spacing) by introducing improved actions and operators,
in order to be able to perform the simulations with larger masses.

Simulations in the HQET, provide the results in the infinite mass
limit, and in principle, also allow for a systematic program of
evaluating corrections to this limit. The corresponding results can be
used as a check of the extrapolations of the results obtained with
propagating heavy quarks, as described above. Since the mass of the
heavy quark does not enter directly into simulations in the HQET
(having been scaled out), lattice artefacts are, a priori, expected to
be as small as for quantities in light quark physics.

\subsection{The Heavy Quark Effective Theory on the Lattice}
\label{subsec:hqetlatt}

The Lagrangian for the HQET, at zero three-velocity (i.e. in the
static limit) in Euclidean space takes the form~\footnote{The
  construction of the HQET in Euclidean space, with $\vec v\neq 0$ is
  subtle, requiring an interpretation of $\delta^{(3)}(\vec x - \vec v
  t)$, see refs.~\protect\cite{mandula,aglietti}}:
\be 
{\cal L} = \bar
hD_4\frac{1+\gamma^4}{2} h\ .
\label{eq:hqetlattice}\ee
If we define the lattice covariant derivative by
\be
D_4\,f(\vec x,t) = \frac{1}{a}\,\left[
f(\vec x,t)-U^\dagger_4(\vec x, t-a)f(\vec x, t-a)\right]\ ,
\label{eq:d4def}\ee
then the propagator of the heavy quark $h$ in the background field 
configuration $\{U\}$ is given by
\be
S^{\{U\}}_h(\vec x,t;\vec 0,0) = \delta^{(3)}(\vec x)\theta(t)
U^\dagger_4(\vec 0,t-a)U^\dagger_4(\vec 0,t-2a)\cdots U^\dagger_4(\vec 0,0)\ ,
\label{eq:hqprop}\ee
with $S_h^{\{U\}}(\vec 0, 0;\vec 0,0)=1$. Other choices of the lattice
covariant derivative correspond to propagators with diffierent
boundary conditions at the origin. From eq.(\ref{eq:hqprop}) we see
that the propagator of the heavy quark is given directly in terms of
the link variables, and so no inversion of a sparse matrix is
necessary (as is the case for the quark propagator in QCD, see
subsection~\ref{subsec:pionprop} above).

\subsection{Calculation of $f_B$ in the Static Limit}
\label{subsec:fbstat}

\begin{figure}
\begin{center}
\begin{picture}(300,60)(0,-25)
\Line(0,0)(60,0)
\Line(0,-2)(60,-2)
\CArc(120,-68.08)(90,48.19,131.89)
\DashCArc(120,66.08)(90,228.19,311.81){4}
\Gluon(180,-1)(240,-1){3}{7}
\Line(240,-1)(280,15)\Line(240,-1)(280,-17)
\BCirc(60,-1){5}
\Text(30,5)[b]{B-Meson}
\Text(280,-1)[l]{Leptons}
\Text(210,8)[b]{W-Boson}
\LongArrow(119,21.92)(121,21.92)
\LongArrow(121,-23.92)(119,-23.92)
\LongArrow(29,-1)(31,-1)
\Text(120,25)[b]{b-quark}
\Text(120,-27)[t]{light antiquark}
\end{picture}
\caption{Quark flow diagram representing the decay of a $B$-meson into 
leptons.}
\label{fig:bdecay}
\end{center}
\end{figure}
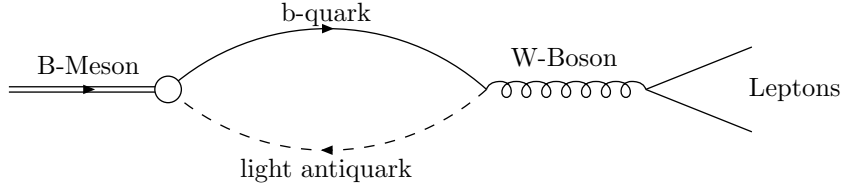

The non-perturbative strong interaction effects in the leptonic decay
amplitude of the $B$-meson are contained in a single parameter, the
decay constant $f_B$. The process is illustrated in the diagram of
fig.~\ref{fig:bdecay}.  To determine this parameter, in analogy to the
computation of $f_\pi$ discussed in subsection~\ref{subsec:pionprop}
above, we compute the correlation function:
\beqn
C_2(t) & = & \sum_{\vec x}\, \<\,0\,|\, \bar h(x)\gamma^4\gamma^5q(x)\ 
\bar q(0)\gamma^4\gamma^5h(0)\,|\,0\,\>\label{eq:c2hqet1}\\ 
& = & -\sum_{\vec x} \, \<\, {\mathrm Tr}[\gamma^4\gamma^5S_l(x,0)
\gamma^4\gamma^5S_h(0,x)]\,\>\label{eq:c2hqet2}\\ 
& = & \frac{|\,\<\,0\,|\bar h\gamma^4\gamma^5q\,|\,0\,\>\,|^2}
{2M_B}\,e^{-{\cal E}t}+\cdots\label{eq:c2hqet3}\\ 
& = & \left(\frac{1}{Z_A^{{\mathrm stat}}}\right)^2\frac{f_B^2M_B}{2}
\ e^{-{\cal E}t} +\cdots \ ,
\label{eq:c2hqet4}\eeqn
where $S_l$ and $S_h$ are the light and heavy quark propagators respectively.
I now explain the features of the calculation represented by 
eqs.(\ref{eq:c2hqet1})--(\ref{eq:c2hqet4}) in some more detail:

\paragraph{What is the interpretation of $\cale$?:}
$\cale$ itself is not a physical quantity, indeed it diverges linearly
with the inverse lattice spacing. At the tree level, it is just the
difference of the mass of the pseudoscalar meson and that of the
quark. When higher order corrections are included the interpretation
of $\cale$ becomes more subtle, and will be explained in the third
lecture below. In the lattice theory, with the hard ultraviolet
cut-off $a^{-1}$, quantum corrections generate a mass term in the
effective theory, even if the bare action does not have such a term
(as in eq.\ref{eq:hqetlattice})\,). This mass term is of
$O(\alpha_s/a)$.  For now I will just note that from the measured
value of $\cale$, it is possible to derive the value of any
short-distance mass of the heavy quark (such as the $\MSbar$ mass).

\paragraph{Heavy Quark Scaling Law:}
Since the action of the HQET is independent of the mass of the heavy
quark, so is the correlation function (\ref{eq:c2hqet1}).  Thus, from
eq.(\ref{eq:c2hqet4}) we can deduce the well known scaling law that
the decay constant of a heavy pseudoscalar meson decreases like the
square-root of its mass. This scaling law is modified by logarithmic
corrections contained in the matching factor $Z_A^{{\mathrm stat}}$ as
explained below.

\paragraph{``Smearing'':}
In practice it is difficult to isolate the contribution of the ground
state in these calculations \cite{eichtencapri,boucaud} using local
interpolating operators.  To enhance the relative contribution of the
lightest state, various extended (or ``smeared'') interpolating
operators are used.  For example one might smear the field in the
axial current by
\be
h^s(\vec x,t)\equiv\sum_{\vec y}f(\vec x,\vec y)\,h(\vec y,t)
\label{eq:hsmeared}\ee
in the Coulomb gauge say. Different choices of the smearing function
$f$ are used by the various groups carrying out such simulations.  The
Fermilab group in particular have been stressing the possibility of
learning about the wave function of the heavy meson by studying its
overlap with different extended operators \cite{eichtensmeared}.
However, it should be remembered that $f_B$ is obtained from the
matrix element of the local (i.e. not smeared) axial current. To
determine this, one has to compute the correlation functions of two
smeared currents (from which one obtains the matrix element of the
smeared current as in eq.(\ref{eq:c2hqet3})) and of a smeared and a
local current (from which one is then able to deduce the matrix
element of the local current). A strong check that the ground state
has been isolated is provided by verifying that the matrix element of
the local current is independent of the smearing function $f$ used
in the calculation

\paragraph{Matching:}
The quantity which one obtains directly in lattice simulations is the
matrix element of the bare axial current in the HQET, defined with the
lattice regularization. In order to obtain $f_B$ one needs the matrix
element of the axial current in QCD. The difference between the two is
an ultraviolet effect and can be calculated in perturbation theory.
Although not necessay, this calculation is usually performed in two
steps 
\be 
A_\mu^{{\mathrm latt,\ HQET}}(a)\to A_\mu^{\overline{{\mathrm MS}},\ 
  {\mathrm HQET}}(\mu) \to A_\mu^{{\mathrm QCD}}\ .
\label{eq:mathching}\ee In the first of these steps lattice
perturbation theory is used to obtain the matrix element of the axial
current in the HQET defined in some continuum renormalization scheme,
such as the $\MSbar$ scheme, from the matrix element of the bare
current in the lattice theory.  In the second step the matrix element
of the physical (QCD) current (and hence $f_B$) is obtained from that
in the effective theory. $Z_A^{{\mathrm stat}}$ in
eq.(\ref{eq:c2hqet4}) is the combined matching factor. It depends
logarithmically on the lattice spacing, being proportional to
$\alpha_s(a)^{(-2/\beta_0)}$, where $\beta_0=11-2/3 N_f$ is the
coefficient of the first term in the $\beta$-function, and $N_f$ is
the number of light-quark flavours.

In these lectures it is not possible for me to describe the technical
details of lattice perturbation theory. The lattice action for the
HQET in eq.(\ref{eq:hqetlattice}) and the QCD action in
eq.(\ref{eq:wilsonaction}), with the derivatives defined as
differences, lead to Feynman rules where the momentum dependence is
given in terms of trigonometric functions. For example the quark
propagator for the Wilson action is given by the expression in
eq.(\ref{eq:sfreew}).  This makes the analytic evaluation of Feynman
diagrams in lattice perturbation theory prohibitatively difficult in
general, and instead the integrals are evaluated numerically.

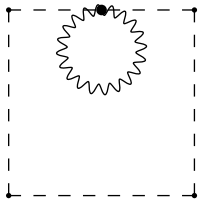
\begin{figure}[t]
\begin{center}
\begin{picture}(80,60)(-5,-5)
\DashLine(0,0)(70,0){5}
\DashLine(70,0)(70,70){5}
\DashLine(0,0)(0,70){5}
\DashLine(0,70)(70,70){5}
\PhotonArc(35,55)(15,0,360){2}{20}
\Vertex(35,70){2}
\Vertex(0,0){1}
\Vertex(70,0){1}
\Vertex(0,70){1}
\Vertex(70,70){1}
\end{picture}
\caption{Representation of a one loop tadpole diagram contributing to
the plaquette variable. The diagram is generated by expanding one of the
link variables to $O(g_0^2)$.}
\label{fig:plaquette}
\end{center}
\end{figure}

Lepage and Mackenzie have argued convincingly that the bare lattice
coupling $g_0^2(a)$ is a poor expansion parameter for lattice
perturbation theory, i.e. that with this choice the higher order
corrections have unnecessarily large coefficients\,\cite{lm}. As an
example, consider the one-loop tadpole correction to the plaquette
variable, illustrated in fig.~\ref{fig:plaquette}.  Numerically it
turns out that such tadpole contributions, which are common to many
quantities, are large. In particular they will appear each time that
there is a link variable (from which the tadpole can emanate). Let
\be
u_0^4\equiv \frac{1}{3}\<\, {\mathrm Tr}\,{\cal P_{\mu,\nu}}\,\>
\label{eq:u0def}\ee
where the plaquette ${\cal P}$ has been defined in
eq.(\ref{eq:plaquettedef}). Thus one might expect that at least part
of the large higher order corrections are given by $u_0$ for each link
variable present in the quantity being studied. If this is the case
then we can replace each link variable $U_\mu$ by $U_\mu/u_0$,
compensating for this by different perturbative matching factors. For
the gluon action one would rewrite ${\cal P}/g_0^2\to {\cal
  P}/(\widetilde g^2 u^4_0)$, with $\widetilde g^2 = g_0^2/u_0^4$.  It
might therefore be expected that $\widetilde g^2$ is a better
expansion parameter, and this appear to be the case (for numerical
evidence of this feature, and of a general discussion of ``tadpole
resummation'' see ref.~\cite{lm}). It would, nevetheless, be very
welcome to have a non-pertubative determination of the matching
factor. For simulations with the Wilson action for the light quark,
$Z_A^{{\mathrm eff}}\simeq 0.7$ for typical beta values used in
current simulations, and for the ``improved'' SW-action described in
subsection~\ref{subsec:propagating}, $Z_A^{{\mathrm eff}}\simeq 0.8$.
These values are obtained from one-loop perturbation theory.

\subsection{Results}
\label{subsec:results}

Before presenting some results obtained from simulations in the heavy quark
effective theory, let us recall precisely what it is that we are
calculating. The decay constant ($f_P$) of a heavy pseudoscalar meson ($P$)
satisfies the
scaling law:
\be
f_P = \frac{A}{\sqrt{M_P}}
\left\{\, \alpha_s^{-2/\beta_0}(M_P)\,(1 + O(\alpha_s)\,)\,+\, O(\frac{1}{M_P})
\right\}\ ,
\label{eq:scaling}\ee
where $M_P$ is the mass of the meson. The factor of
$\alpha_s(M_P)^{-2/\beta_0}$ arises from the matching of the HQET with
QCD, and in our case above (for the $B$-meson) it is included in
$Z_A^{{\mathrm stat}}$. From lattice simulations in the HQET, we
obtain $A$, the normalisation of the leading term. In order to compute
the $O(1/M_P)$ corrections, it is necessary to compute the matrix
elements of the dimension-4 operators which arise in the expansion of
the axial current in QCD in terms of operators of the HQET. It is also
necessary to include the $O(1/M_P)$ corrections to the heavy quark
action (\ref{eq:hqetlattice}). Such calculations are only just
beginning to be performed~\cite{fb1overm}. I denote by $f_B^{{\mathrm
    stat}}$, the value of $f_B$ obtained by dropping the $1/M_P$
corrections. The UKQCD collaboration obtain
\beq
f_B^{{\mathrm stat}} = 266\err{18}{20}({\mathrm stat})\err{28}{27}
({\mathrm syst})\,{\mathrm MeV}\ ,
\eeq
where the quoted systematic error is due to the uncertainty in the value
of the lattice spacing (which enters as $a^{3/2}$ in this calculation).
For a recent review of the status of computations of $f_B^{{\mathrm stat}}$
see ref.~\cite{cramelbourne}. I will delay the interpretation of this result
until after a discussion of the computations performed with propagating
heavy quarks in subsection~\ref{subsec:propagating}.

The calculation of $f_B^{{\mathrm stat}}$ is performed in a theory
without any explicit large masses. A priori, one would therefore
expect that the systematic errors should be similar to those in light
quark physics. By repeating this calculation at several values of
$\beta$, the Fermilab group have claimed that there are large lattice
artefacts~\cite{fnal}.  This conclusion relies heavily on the results
at $\beta=5.7$ however, and a detailed study of the data of the APE
collaboration leads to the conclusion that the results for
$f_B^{{\mathrm stat}}$ are independent of $\beta$ (i.e. of $a$) for
$\beta \ge 6.0$~\cite{apefbstat}.

\subsection{Calculations of $f_B$ performed with Propagating Heavy Quarks}
\label{subsec:propagating}

In this subsection I return to simulations performed in QCD. Now,
as explained above, 
 the $b$-quark
is too heavy to use, so the calculations of decay constants and other
physical quantities are performed with quark masses in the region of
that of the charmed quark. Even so, the largest source of systematic
uncertainty are discretisation errors which are of $O(m_Q a)$ for a
heavy quark with mass $m_Q$. For example, at $\beta = 6.2$ the bare mass
of the charm quark, $m_{ch,0}$, is about 0.35 in lattice units, where
\be
m_{ch,0} \, a \equiv
\frac{1}{2}\left(\frac{1}{\kappa_{ch}}-\frac{1}{\kappa_c}\right) \,
\label{eq:mc}\ee
and $\kappa_{ch}$ is the $\kappa$-parameter corresponding to a charm
quark.

\paragraph{``Improvement'':}
In order to reduce these discretisation errors much effort is being
invested to develop the systematic ``improvement'' technique of
Symanzik~\cite{symanzik}. This technique involves the modification of
the lattice action and operators in such a way that the discretisation
errors are formally reduced. As an example consider the
action~\cite{sii}:
\be
S^{II}_{QCD} = S_{\mathrm gluon} + S_W^F -\Delta S^{II}
\label{eq:sii}\ee
where
$S_{\mathrm gluon}$ and $S_W^F$ are the Wilson gluon and quark actions as defined
in subsection~\ref{subsec:latticeqcd} and 
\beqn
\lefteqn{\Delta S^{II} = -
2\kappa a^4\,\sum_{x,\mu}\frac{r}{8a}\Big\{
\bar\psi(x)U_\mu(x)U_\mu(x+a\hat\mu)\psi(x+2a\hat\mu) + }\nn\\ 
& & \bar\psi(x+2a\hat\mu)U^\dagger_\mu(x+a\hat\mu)U^\dagger_\mu(x)\psi(x)
-2\bar\psi(x)\psi(x)\Big\} \ .
\label{eq:deltasii}\eeqn
Now expanding $S^{II}_{QCD}$ as a series in the lattice spacing we find:
\be
S_{QCD}^{{\mathrm cont}}= S_{QCD}^{II} + O(a^2)\ ,
\label{eq:sexp}\ee
where the superscript ``cont'' represents ``continuum''. A similar
expansion for the Wilson action,
eq.(\ref{eq:oa}), gives discretisation errors of $O(a)$
rather than of $O(a^2)$. When radiative corrections are included one
finds that the discretisation errors are of $O(\alpha_s
a)$~\cite{heatlie}, and so the use of the action (\ref{eq:sii}) has
formally reduced the lattice artefacts from $O(a)$ to $O(\alpha_s a)$.
One can also readily check that the problem of fermion doubling has
not been reintroduced by the addition of the term $\Delta S^{II}$.

The action in eq.(\ref{eq:sii}) contains next-to-nearest neighbour
interactions, in addition to the nearest-neighbour interactions of the
Wilson theory.  By changing (the fermionic) variables in the functional
integral, it is also possible to construct an equivalent improved
theory with nearest neighbour interactions only~\cite{sw}
\be
S^{I}_{QCD} = S_{\mathrm gluon} + S_W^F -\Delta S^{I}
\label{eq:si}\ee
with
\be
\Delta S^I = \frac{ira\kappa}{2}g_0
\sum_{x,\mu\nu}\bar\psi(x)\sigma_{\mu\nu}F_{\mu\nu}\psi(x)\ .
\label{eq:deltasi}\ee
This lattice action, which was first proposed by Sheihkoleslami and
Wohlert, is called the SW-action.  In eq.~(\ref{eq:deltasi}),
$F_{\mu\nu}$ is a lattice definition of the field strength tensor.
Several groups performing large scale numerical simulations are using
this action, and the results from the UKQCD collaboration which I am
using to illustrate this talk were obtained with this formulation of
QCD.  Even with this action however, in simulations containing the
charm quark, one might expect 10\% errors due to finite-$a$ effects.

It is also possible to try to construct an action by subtracting
$\Delta S^I$ with a coefficient which is different from unity in order
to reduce the lattice artefacts still further. This is sometimes done
at the one-loop level~\cite{naik} or by using mean field theory
estimates~\cite{fnalhyperfine,km} or
non-perturbatively~\cite{nonpertren2}.  Finally let me mention the
exciting program of constructing ``Perfect Actions''~\cite{perfect},
in which, ideally, by using the renormalization group transformations
on a fixed-point action, all the discretisation errors would be
removed (although in practice one may have to settle for a substantial
reduction of the artefacts, rather than for ``perfection'').

\begin{figure}[t]
  \centerline{\hspace{15cm}\epsfxsize=8cm\epsffile{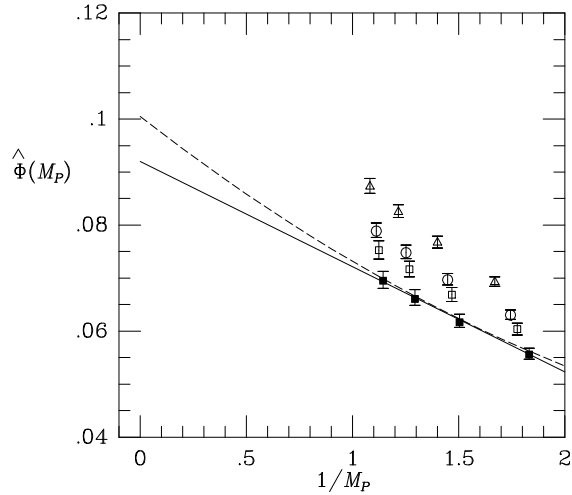}}
  \caption{\label{fig:extrap} The behaviour
    of the scaling quantity $\hat \Phi$ with $1/M_P$. The solid line
    represents the linear fit to the chirally extrapolated points
    using the three heaviest meson masses, whereas the dashed curve
    results from a quadratic fit to all four.}
\end{figure}

\paragraph{Results:}
My summary of the results obtained with propagating quarks is: 
\beqn
f_D & = & 200\pm 30\, {\mathrm MeV}\label{eq:fdresult}\\
f_B & = & 180\pm 40 \,{\mathrm MeV}
\label{eq:fbresult}\ , \eeqn 
where the result for the $B$-meson was obtained, of course, by
extrapolation to the $b$-quark of results obtained with lighter
quarks. This extrapolation is illustrated in 
fig.~\ref{fig:extrap}, where the
results from the UKQCD collaboration are presented for the scaling
quantity 
\be
\hat\Phi(M_P)\equiv f_P\, M_P^{1/2}\, 
(\alpha_s(M_P)/ \alpha_s(M_B))^{2/\beta_0}
\, , 
\label{eq:phihatdef}\ee
as a function of $1/M_P$~\cite{quenchedfd}, where all quantities are
given in lattice units~\footnote{The simulation was performed at
  $\beta=6.2$ for which the inverse lattice spacing is about 2.9
  GeV.}.  Here $P$ represents a heavy-light pseudoscalar meson. There
are twelve ``measured'' points, corresponding to three light quark
masses (represented by the open triangle (heaviest), circle and
square(lightest)) and four heavy quark masses (represented, for each
light quark mass, by the same open symbol).  The full points represent
the values obtained by extrapolation of the light-quark masses to the
chiral limit, for each heavy quark mass. If there were no $1/M_P$ and
higher corrections then, for each choice of the light quark mass, the
points would lie on a horizontal line.  Clearly there are significant,
negative, $O(1/M_P)$ corrections. The violation of the heavy quark
scaling law is of the correct sign to match up with the static result,
although UKQCD find a (small) discrepancy between their result for
$f_B^{{\mathrm stat}}$ and the value obtained by extrapolating the
results computed at finite values of $M_P$. Given the different
systematic errors in the computations with static and with propagating
heavy quarks, the near consistency of the results is very pleasing. I would
summarize the best estimate of $F_B$ from lattice simulations as being
190(40) MeV.

\begin{figure}[t]
  \centerline{\hspace{15cm}\epsfxsize=8cm\epsffile{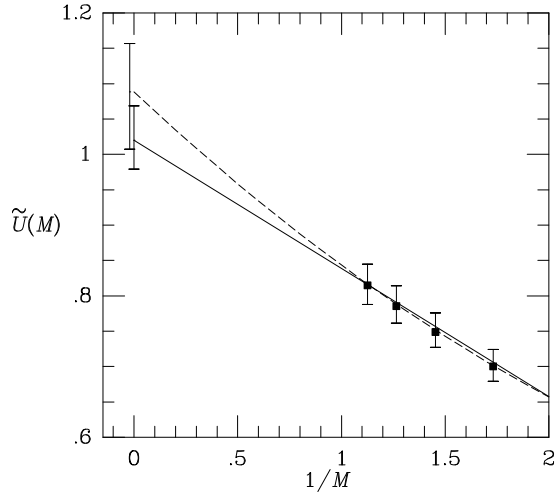}}
  \label{fig:test}
  \caption{The quantity
        $\widetilde U(M)$ plotted against the inverse spin-averaged
        mass. Linear and quadratic fits are represented by the solid
        and dashed curves respectively. Also shown are the statistical
        errors of the extrapolation to the infinite mass limit.}
\end{figure}

\paragraph{Test of the Heavy Quark Symmetry:}
I would like to mention one important test of the lattice calculations
with propagating heavy quarks, which is provided by the heavy quark
symmetry~\cite{abada}. In the infinite mass limit the decay constants
of the pseudoscalar ($f_P$) and vector ($f_V$) mesons are related
\be
\widetilde U(M)\equiv\frac{f_Vf_P}{M}
\frac{1}{1+\frac{8}{3}\frac{\alpha_s(M)}{\pi}}\to 1
\label{eq:test}\ee
as $M\to\infty$, where $f_V$ is defined by
\be
\<\,0\,|V_\mu(0)|\,V\,\> = \epsilon_\mu\,\frac{M_V^2}{f_V}\ .
\label{eq:fvdef}\ee
In eq.(\ref{eq:fvdef}), $V_\mu$ is the vector current with the
appropriate flavour quantum numbers, and $\epsilon_\mu$ is the
polarisation vector of the vector meson $V$. The perturbative matching
factor in eq.(\ref{eq:test}), has been calculated to one-loop order.
In fig.~\ref{fig:test} I present the results of the ratio in
eq.(\ref{eq:test}) as a function of $1/M$, where for $M$ I have taken
the spin averaged meson mass ($M=(3\,M_V+M_P)/4$)~\cite{quenchedfd}.
It can be seen that the results obtained at finite values of the mass
of the heavy quark differ substantially from the asymptotic value, but
the extrapolation to $M=\infty$ is in good agreement.

\subsection{Other Physical Quantities in $B$-Physics}
\label{subsec:other}

To conclude this lecture I will briefly mention some other quantities
in heavy quark physics which have been, or are being, computed in
simulations in the HQET. 

\paragraph{${\mathbf f_{B_s}/f_{B_d}}$:}
By studying the dependence on the results for $f_B$ on the mass of the
light quark, one obtains a good estimate of the ratio of the decay
constants of the $B_s$ and $B_d$-mesons. With both static and
propagating heavy quarks one typically finds that $f_{B_s}$ is about
10-20\% larger than $f_{B_d}$. This can be understood as being due to
the fact that the $B_s$ meson is more compact than $B_d$, with a
larger wave-function at the origin.

\paragraph{Heavy Baryons:}
The detailed study of heavy baryons is just beginning in lattice
simulations. In the HQET it is possible to evaluate mass differences,
such as that between the $\Lambda_b$ baryon and $B$-meson. The UKQCD
collaboration find for this quantity~\cite{static}: 
\be
M_{\Lambda_b}-M_B = 420\err{100}{90}\err{30}{30}\ {\mathrm MeV}\ ,
\label{eq:baryonhqet}\ee
to be compared with the experimental result of 346 MeV with an error
of about 7 MeV (this is based on the new result from the CDF
collaboration~\cite{cdf}). It must be remembered that the result in
eq.(\ref{eq:baryonhqet}) represents the value in the infinite mass
limit.  A comprehensive study of the spectrum of heavy baryons using
propagating heavy quarks, can be found in ref.~\cite{baryons}, where
the $\Lambda_b$-$B$ mass splitting was found to be $359\pm 50\pm 27$\,MeV.

\paragraph{${\mathbf B_B}$:}
One of the most important quantities in the phenomenology of $B$-physics 
is the parameter $B_B$, which together with $f_B$ contains the strong
interaction effects in $B^0-\overline{B}^0$ mixing. It is defined by
\beq
B_B=[\alpha_s(\mu)]^{-2/\beta_0}\,\frac{\<\overline {B}^0\,|O_L(\mu)
|B^0\,\>}{\frac{8}{3}f_B^2M_B^2}
\label{eq:bbdef}\ee
where $O_L$ is the $\Delta B=2$ operator
\be
O_L(\mu)=\bar b\gamma_\mu(1-\gamma^5)q\,\bar b\gamma^\mu(1-\gamma^5)q\ .
\label{eq:oldef}\ee
The first factor on the right hand side of eq.(\ref{eq:bbdef}) is introduced
to cancel the $\mu$ (i.e. renormalization scale) dependence of the matrix
element at leading logarithmic order.

In order to determine the physical amplitude for the $\Delta B=2$ transition,
it is necessary to write the corresponding QCD operator, renormalized
in some continuum scheme, in terms of the bare operators of the HQET
defined using the lattice regularization. Because of the breaking of
chiral symmetry in theories based on the Wilson formulation of lattice
fermions, this requires the evaluation of the matrix elements of 3 other
$\Delta B=2$ operators in addition to $O_L$,
\beqn
O_R & = & \left(\bar b \gamma_\mu(1 + \gamma^5) q\right)
\left(\bar b \gamma_\mu(1 + \gamma^5) q\right)\label{eq:ordef}\\ 
O_N & = & \left(\bar b \gamma_\mu(1 - \gamma^5) q\right)
\left(\bar b \gamma_\mu(1 + \gamma^5) q\right) + \nn\\ 
& & \mbox{}2\left(\bar b (1 - \gamma^5) q\right)
\left(\bar b (1 + \gamma^5) q\right)
+2\left(\bar b (1 + \gamma^5) q\right)
\left(\bar b (1 - \gamma^5) q\right)\label{eq:ondef}\\ 
O_S & = & \left(\bar b (1 - \gamma^5) q\right)
\left(\bar b (1 - \gamma^5) q\right)\label{eq:osdef}\ ,
\eeqn
where the QCD operator $O_L$ is given by
\be
O_L^{QCD}(m_b) = Z_L\,O_L^{\rm latt} + Z_R\,O_R^{\rm latt} + 
Z_N\,O_N^{\rm latt} + Z_S\,O_S^{\rm latt} \ .
\label{eq:bparzs}\ee
The matching coefficients $Z_i,\ \{i=L,R,N,S\}$, have been calculated at
one loop level for the Wilson and SW actions in
references \cite{fhh} and \cite{bp} respectively. The one-loop
contribution to $Z_L$ is large, for example at $\beta = 6.2$ with the
SW action (corresponding to the UKQCD simulation), $Z_L\simeq 0.55$.
This underlines further the necessity for a non-perturbative determination
of the matching coefficients.

\begin{figure}[t]
\begin{center}
\begin{picture}(300,180)(-130,-120)
\Line(-110,0)(110,0)\Line(-110,-2)(110,-2)
\CArc(-55,0)(55,0,180)\CArc(55,0)(55,0,180)
\Vertex(-110,-1){1.5}\Vertex(0,-1){1.5}\Vertex(110,-1){1.5}
\Text(-110,-5)[t]{\bf S}\Text(110,-5)[t]{\bf S}\Text(0,-5)[t]{\bf L}
\Line(-145,-20)(145,-20)
\Line(-110,-85)(-4,-85)\Line(4,-85)(110,-85)
\Line(-110,-87)(-4,-87)\Line(4,-87)(110,-87)
\Text(-135,-65)[c]{\Large $\frac{8}{3} $}
\Vertex(-110,-86){1.5}\Vertex(-4,-86){1.5}
\Vertex(4,-86){1.5}\Vertex(110,-86){1.5}
\CArc(-57,-85)(53,0,180)\CArc(57,-85)(53,0,180)
\Text(-110,-90)[t]{\bf S}\Text(110,-90)[t]{\bf S}
\Text(-4,-90)[tr]{\bf L}\Text(4,-90)[tl]{\bf L}

\Text(-110,-110)[c]{$-t_1$}\Text(0,-110)[c]{$0$}\Text(110,-110)[c]{$t_2$}
\LongArrow(145,-110)(175,-110)\Text(160,-105)[b]{time}
\end{picture}
\caption{Diagramatic representation of the ratio defined in eq.(\protect
  \ref{eq:ridef}). The double (single) lines represent the propagators
  of the heavy (light) quark. The labels $S$ and $L$ indicate whether
  smeared or local operators have been inserted at the corresponding
  point.}
\label{fig:ratio}
\end{center}
\end{figure}
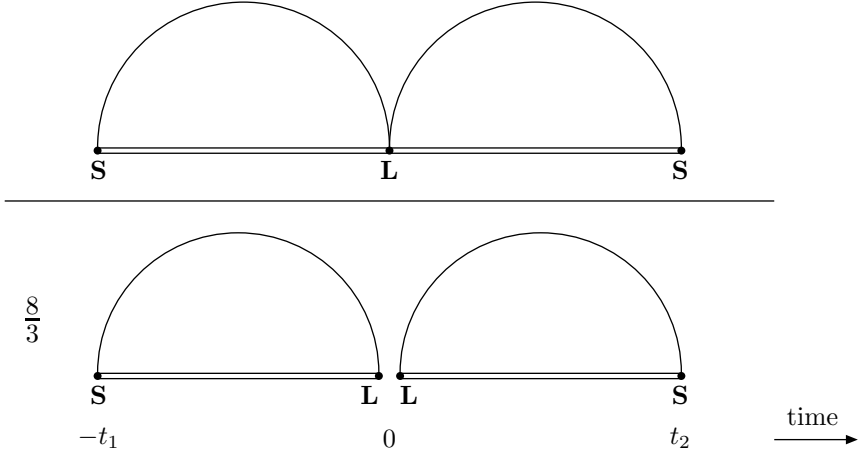

The matrix elements of the lattice four-quark operators, normalised by
the factor $8/3 f_B^2m_B^2$ as in eq.(\ref{eq:bbdef}), can readily be
obtained from the ratio of correlation functions:
\beq
R_i^{SS}(t_1,t_2) = \frac{K_i^{SS}(t_1,t_2)}
{\frac{8}{3}C^{SL}(t_1)C^{SL}(t_2)}
\label{eq:ridef}\ee
where $K_i^{SS}$, with $i=L,R,N,S$, is the three point function
\beqn
K_i^{SS}(t_1, t_2) & \equiv &
\sum_{\vec x_1,\vec x_2}\,\<\,0\,|\left\{A^{\dagger^S}(\vec x_1, -t_1)\,
O_i^{{\rm latt}}(0)\,A^{\dagger^S}(\vec x_2, t_2)\right\}
 \,|\,0\,\>\label{eq:kdef}\\ 
&\stackrel{t_1,t_2 \gg 0}{\to} &
\frac{(Z^S)^2}{2 M_B}e^{-E(t_1+t_2)}
\<\overline{B}^0\,|O_i^{{\rm latt}}\,|B^0\>\ ;
\label{eq:kasymp}\eeqn
$C^{SL}$ is the two-point correlation function of two axial currents,
\be
C^{LS}(t) = \sum_{\vec x}
\,\<\,0\,|A_4^L(\vec x, t)\,A_4^{\dagger^S}(\vec 0, 0)\,|0\,\>\ ,
\label{eq:clsdef}\ee
and the superscripts $L$ and $S$ stand for {\it local} and {\it smeared}
respectively. The smeared currents are introduced to enhance the 
overlap with the ground state, and the local ones because it is from these
that we are able to obtain the decay constant $f_B$. The ratio $R_i$
is represented by the diagram in fig.\ref{fig:ratio}.

The UKQCD collaboration finds the following results from their
simulations at $\beta = 6.2~$\cite{static}: 
\beqn B_{B_d} & = & 1.02\err{5}{6}\,{\mathrm (stat)}\, \err{3}{2}\, {\mathrm
  (syst)}\label{eq:bbdresult}\\
B_{B_s} & = & 1.04\err{4}{5}\,{\mathrm (stat)}\, \err{2}{1}\, {\mathrm
  (syst)}
\label{eq:bbsresult} \eeqn 
A discussion of the phenomenological consequences of these results can
be found in ref.\cite{static}. A preliminary result which is somewhat
higher (by about 25\%) than that in eqs.(\ref{eq:bbdresult}) and
(\ref{eq:bbsresult}) was recently presented in ref.~\cite{draper}.

Relatively few calculations have been performed for the $B$-parameter
with propagating heavy quarks~\cite{bs,abada,jlqcd}, the most recent
of which quotes 
\be
\alpha_s^{(2/\beta_0)}(5\,{\mathrm GeV})\ B_B =
0.90(5)\ {\mathrm and}\ 0.84(6)
\label{eq:bbpropagating}\ee
obtained using the Wilson fermion action at $\beta = 6.1$ and 6.3
respectively~\cite{jlqcd}~\footnote{Thus the JLQCD group quote their
  results in terms of the matrix element renormalized at 5 GeV in the
  $\MSbar$ scheme, and not the renormalization group improved
  one~\cite{jlqcd}.}. Thus this result obtained with propagating
Wilson quarks is a little larger (about 20\% larger) than that
obtained with static quarks in eq.(\ref{eq:bbdresult}). It is too
early to be sure that this difference is a physical effect, rather
than an artefact of lattice systematics. In particular it will be
interesting to compare the calculations of the matching factors in the
recent studies~\cite{draper,jlqcd} when the details of these
calculations are published, with those of ref.~\cite{static,abada}.
In reference~\cite{abada} there is some evidence that the $O(1/m_Q)$
corrections to the result obtained in the static limit may be negative,
so that the $B$-parameter increases slowly as the mass of the heavy
quark increases, which is the opposite conclusion one would draw from
comparing the results in eqs.(\ref{eq:bbdresult}) and
(\ref{eq:bbsresult}), with that of the JLQCD collaboration with
propagating quarks.

\paragraph{The Hyperfine Splitting, $M_{B^*}-M_B$:}
As a final example of quantities which can be evaluated in lattice
simulations, consider the mass difference between the vector and
pseudoscalar heavy-light mesons. In the static limit these mesons are
degenerate, and their mass difference is of $O(1/m_Q)$, where $m_Q$ is
the mass of the heavy quark.  The hyperfine splitting is given by the
matrix element of the chromomagnetic operator which appears at
$O(1/m_Q)$ in the action. The results of lattice calculations of the
hyperfine splitting give values which are considerably lower (by a
factor of 2 or so) than the experimental 
result~\cite{bocicchio,static,gms}. 
For example in ref.~\cite{static} it was found that
\be
M_{B^*}^2-M_B^2 = 0.281\err{15}{16}\,{\mathrm (stat)}\err{40}{37}\,
{\mathrm (syst)}\,{\mathrm GeV}^2\ ,
\label{hyperfineresult}\ee
where the systematic error represents the uncertainty in the value of
the lattice spacing only. The experimental result for this quantity is
$0.488(6)\,$GeV$^2$~\cite{hyperfineexperiment}. 

It has been observed for a considerable time that the values of the
hyperfine splitting in heavy mesons (including heavy quarkonia) are
found to be too small in lattice simulations with propagating heavy
quarks. It is understood that this physical quantity has particularly
large discretisation errors, and indeed as one ``improves'' the
action, the discrepency between the lattice results and experimental
measurement decreases~\cite{ukqcdhyperfine,fnalhyperfine}.  It is
peculiar however that the result obtained with the HQET is so
different from the experimental value. In the HQET the discretisation
errors are all of $O(a\lqcd)$, since there is no dependence on the
mass of the heavy quark. Of course quenching may be partly responsible
for the discrepency, and in addition the matching factor (between the
lattice operator in the HQET and that in QCD) is found to have very
large one-loop corrections (the factor, which is 1 at the tree level,
is found to be about 1.8 at the one-loop level). This is another
example of the necessity of computing these matching factors
non-perturbatively.

\section{Lecture 3 -- Renormalons and Power Divergences in Lattice Simulations}
\label{sec:renormalons}

In the final lecture I will briefly review the status of a subject
which is currently very topical, not just in lattice simulations, but
in phenomenological studies of QCD in general. This is the problem of
renormalons~\cite{renormalons}, which arises when one is interested in
computing power corrections to hard scattering and decay processes.
For example, in the discussion above this problem would arise if one
were to compute the $O(1/m_Q)$ corrections to the decay constant of a
heavy meson.  Other important examples include the evaluation of
higher twist effects in deep-inelastic structure functions and the
contributions of the gluon condensate to quantities which are measured
in $e^+$-$e^-$ annihilation. The discussion in this lecture will be
restricted to those processes for which an operator product expansion
exists.

\subsection{Overview}
\label{subsec:overview}

I will now present the principal results concerning the appearance and
cancellation of renormalon singularities in operator product
expansions, using the Heavy Quark Expansion as an illustration.
Consider the evaluation of the matrix element of some local QCD
operator containing one or more heavy quark fields. We call this
operator $\oqcd$. For example if we wish to evaluate the leptonic
decay constant $f_B$ as in section~\ref{subsec:fbstat}, then $\oqcd$
would be the axial current $\bar b \gamma^\mu \gamma^5 q$, where $q$
represents the field of the light quark. Using the HQET we expand
$\oqcd$ in inverse powers of the mass of the heavy quark ($m_Q$):
\beqn
\lefteqn{\oqcd = C_1(m_Q/\mu)\, O_1^{{\rm HQET}}(\mu) \, + }
\nonumber\\ 
& & \frac{1}{m_Q}
C_2(m_Q/\mu)\, O_2^{{\rm HQET}}(\mu) + O(1/m_Q^2)
\label{eq:ope}\eeqn
where $\mu$ is the scale at which the operators of the HQET, $O_1^{{\rm
HQET}}$ and $O_2^{{\rm HQET}}$, are renormalized. We consider here the
simple situation for which there is a single operator in each of the
first two orders of the expansion, but the discussion below can be
easily generalized to the case in which there are more operators. 

When evaluating matrix elements of $\oqcd$ beyond the leading order in
the $1/m_Q$ expansion, in addition to the corresponding higher
dimension operators on the r.h.s.$\!$ of eq.(\ref{eq:ope}), it is also
necessary to keep higher order terms in the heavy quark action, i.e.
by replacing the Dirac term in the QCD action by
\beqn
\lefteqn{\bar Q(i\,\gamma\cdot D -m_Q) Q \rightarrow
\bar h_v(iv \cdot D)h_v + 
\frac{1}{2m_Q}\bar h_v(iD)^2h_v} \nn\\
& & +\frac{c_{{\rm mag}}}{2m_Q}\frac{g}{2} \bar h_v\sigma_{\alpha\beta}
F^{\alpha\beta}h_v + O(1/m_Q^2)\ ,
\label{eq:hqet}\eeqn
where $Q$ and $h_v$ represent the fields of the heavy quark in QCD and
the HQET respectively, and $v$ is the quark's four-velocity. In the
discussion below we will take the quark to be at rest, and will denote
the corresponding heavy quark field by $h$ (without any subscript).  
$c_{{\rm mag}}$ is a
constant determined by matching the effective theory onto the full one
(QCD). The corresponding constant for the kinetic term ($\bar
h(iD)^2h$) is equal to one by reparametrization invariance
\cite{repar,mnalmunecar}.  Throughout the discussion below it is
implied that the action of the HQET contains sufficiently many terms
for the precision of the calculation.

In general the QCD operator $\oqcd$ also requires renormalization and
is defined at some scale $M$. Unless specifically needed I will
suppress the dependence on $M$ in $\oqcd$ and in the coefficient
functions $C_i$.

I will now summarize the main points which I wish to make in this
lecture, a more detailed account can be found in 
ref.~\cite{ms1,melbourne,ms2}:
\begin{itemize}
\item If we restrict the calculation to the leading order in $1/m_Q$,
  i.e. if we neglect the $O(1/m_Q)$ terms, then the perturbation
  series for the coefficient function $C_1$ diverges, and is not Borel
  summable.  This is due to a singularity in the Borel transform of
  the perturbation series, called an {\it infra-red renormalon}. This
  implies that there is an ambiguity in the evaluation of $C_1$,
  coming from the different possible ways of defining the sum of the
  series. This ambiguity is of $O(1/m_Q)$.
\item Formally therefore, we should not include the $O(1/m_Q)$
  corrections in eq.(\ref{eq:ope}) until we have computed sufficiently
  many terms in the perturbation series for $C_1$ to control its
  divergent behaviour.
\item For this talk I will restrict the discussion to problems for
  which the matrix elements of $\oqcd$ have no renormalon ambiguities.
  In general such non-perturbative effects do exist, and appear on the
  right hand side of eq.(\ref{eq:ope}) in the matrix elements of the
  operators of the HQET. They do not affect the coefficient functions.
\item In renormalization schemes based on the dimensional
  regularization of ultraviolet divergences in the HQET, such as the
  \msbar\ renormalization scheme, the matrix elements of $O_2^{{\rm
      HQET}}$ are also not Borel summable in perturbation theory, due
  to an {\it ultraviolet renormalon} singularity in their Borel
  transforms.  The corresponding ambiguities in the matrix elements of
  $O_2^{{\rm HQET}}$ cancel those in $C_1$. Predictions for physical
  quantities cannot have any ambiguity, of course, but the matrix
  elements of higher-dimensional, or higher-twist, operators are, in
  general, not defined uniquely by the $\MSbar$ procedure.
\item If a hard ultraviolet cut-off is used, such as the lattice
  spacing in the lattice HQET, then the matrix elements of $\otwohqet$
  do not have ambiguities due to ultraviolet renormalons. Indeed the
  matrix elements can be computed (unambiguously) in lattice
  simulations, in contrast to those in the $\MSbar$ scheme, which
  cannot be computed directly using some non-perturbative technique.
  This, in turn, implies that the coefficient functions $C_1$ do not
  contain ambiguities due to infra-red renormalons. In the lattice
  theory it is natural to present the discussion in terms of bare
  operators in the effective theory, defined with the lattice spacing
  $a$ as the ultraviolet cut-off. I will assume here that $m_Q a\gg
  1$. Of course if the inverse lattice spacing was much smaller than
  the heavy quark mass then there would be no need to use the HQET.
\item The absence of renormalon ambiguities in $C_1$ in the lattice
  theory is due to a cancellation between terms which, in any order of
  perturbation theory, are of different order in the $1/m_Q$
  expansion, i.e.  between terms which behave logarithmically with
  $m_Q a$ and non-leading ones of $O(1/m_Qa)$.
\item In the lattice theory, matrix elements of higher dimensional
  operators (such as $\otwohqet$) diverge as inverse powers of the
  lattice spacing, and are hence manifestly unphysical. For example,
  for the matrix element of the kinetic energy operator, \beq \langle
  H|\bar h \vec D^2 h | H \rangle \sim O(1/a^2)
  \label{eq:power}\eeq
  where $H$ represents the heavy hadron. This can readily be confirmed
  in perturbation theory.
\item The subtraction of the power divergences in perturbation theory
  introduces renormalon ambiguities. Thus for example, the
  perturbation series of the terms which diverge quadratically, i.e.
  those which are $O(1/a^2)$, in eq.(\ref{eq:power}) contains a
  renormalon ambiguity.
\item The perturbation series for the pole mass has a renormalon
ambiguity of $O(\lqcd)$ \cite{bsuv,bb}. This is not the case for 
short-distance definitions of the heavy quark mass, such as 
\beq
\overline m_Q \equiv m_Q ^{\overline{{\rm MS}}}
(m_Q ^{\overline{{\rm MS}}})\ .
\label{eq:mbardef}\eeq
\item It is possible to compute $\mbar_Q$ (and other short-distance
  definitions of the mass), using only simulations in the HQET. The
  quantity which is computed directly in lattice simulations is the
  bare binding energy, $\cale$ in eq.(\ref{eq:c2hqet3}). $\cale$ is
  also not a physical quantity, diverging linearly with the inverse
  lattice spacing. It's significance can be deduced by matching QCD
  and the lattice formulation of the HQET, from which one finds
  \begin{equation}
  \cale = M_P - m_{{\mathrm pole}} + \delta m\ ,
  \label{eq:caleint}\end{equation}
where $M_P$ is the mass of the pseudoscalar meson~\footnote{Other
  hadronic states can also be used equally well to determine
  $\mbar$.}, $m_{{\mathrm pole}}$ is the pole mass of the heavy quark,
and $\delta m$ the perturbative series of linearly divergent terms
contributing to the mass renormalization in the effective theory with
the action (\ref{eq:hqetlattice})~\cite{cgms}. The perturbation series
for $m_{{\mathrm pole}}$ in terms of $\mbar$ has a renormalon
singularity, which is cancelled by that in $\delta m$. The linear
divergence in $\delta m$ cancels that in $\cale$. Thus using
eq.(\ref{eq:caleint}), the measured value of $\cale$ and perturbation
theory, one can obtain $\mbar$.  In ref.\cite{cgms} it was found in
this way that
\beq 
\mbarb = 4.17\pm 0.05 \pm 0.03\, {\rm GeV} + O(1/m_b) \ .
\label{eq:mbresult}\eeq
\item The presence of renormalons in physical quantities for which
  there is no Operator Product Expansion, such as the Drell-Yan
  process or in event shape variables in jet physics, is a subject
  currently under intensive investigation \cite{sterman1,az}.
  It is hoped that these studies will provide important
  phenomenological information about the sub-asymptotic (non-leading
  twist) behaviour of physical quantities. In lattice QCD there are
  analogous questions about the presence of ambiguities in
  perturbation series for quantities which contain power divergences,
  e.g. is there an ambiguity in the perturbative evaluation of the
  critical mass when using Wilson fermions?
\end{itemize}

\section*{Final Remarks}

I hope that in these lectures I have managed to convey the exciting
possibilities which lattice simulations offer for the computation of
non-perturbative strong interaction effects in physical quantities.
The evaluation of these quantities is important to progress in
particle physics. The applications to $B$-physics which were discussed
in the secong lecture, are only a small subset of the phenomelogically
relevant quantities which are being computed using ``on the lattice''
(see for example refs.~\cite{latt93,latt94,latt95}).  At the same time
I have tried to explain the difficulties which remain, of which a
reliable formalism and algorithms for the inclusion of light quark
loops is, mot probably, the most significant.

\subsection*{Acknowledgements}

I warmly thank the partcipants of this school, and especially the
organisers, Maria-Jose Herrero and Fernando Cornet, for their interest
and questions, and for creating such a stimulating and enjoyable
atmosphere.

I acknowledge the Particle Physics and Astronomy Research Council for
its support through the award of a Senior Fellowship. I also
acknowledge partial support by the EC contract CHRX-CT92-0051.


\begin{thebibliography}{99}    
\bibitem{latt93} C.W.Bernard, \np {B (Proc.Suppl.) 34} (1994) 47;
C.T.H.Davies, {\it ibid.} 135; R.D.Kenway, {\it ibid.} 153
\bibitem{latt94} R.Gupta, \np {B (Proc.Suppl.) 42} (1995) 85;
G.Martinelli, {\it ibid.} 127; C.Michael, {\it ibid.} 147;
J.H.Sloan, {\it ibid.} 171; R.Sommer, {\it ibid.} 186
\bibitem{latt95} J.N.Simone, Fermilab preprint CONF-96-017-T (1995)\\ 
(hep-lat/9601017); \\ A.Soni, Brookhaven preprint BNL-62284 (1995) 
(hep-lat/9510036)
\bibitem{cramelbourne} C.R.Allton, Rome University preprint
ROME-111 (1995)\\  (hep-lat/9509084)
\bibitem{melbourne} C.T.Sachrajda, University of Southampton preprint,
SHEP-95-32 (1995) (hep-lat/9509085)
\bibitem{mm} I.Montvay and G.Munster, {\it Quantum Fields on a Lattice},
Cambridge University Press, Cambridge (1994) 
\bibitem{creutz} M.Creutz, {\it Quarks, Gluons and Lattices},
Cambridge University Press, Cambridge (1983) 
\bibitem{nn} H.B.Nielsen and M.Ninomija, \np {B 185} (1981) 20 
(erratum \np {B 195} (1982) 541; \np {B 193} (1981) 173
\bibitem{bochicchio} M.Bochicchio et al., \np {B 262} (1985) 331
\bibitem{symanzik} K.Symanzik, in ``Mathematical Problems in Theoretical
Physics'', \\ Springer Lecture Notes in Physics,
vol. 153 (1982) 47, ed. R.Schrader, R.Seiler and D.A.Uhlenbrock
\bibitem{glueballs} C.Michael, Liverpool University Preprint LTH-370
(1996) \\ (hep-ph/9605243)
\bibitem{ukqcdstrange} UKQCD Collaboration, C.R.Allton et al.,
\pr {D 49} (1994) 474
\bibitem{sharpe} S.R.Sharpe, \np {B (Proc.Suppl.) 30} (1993) 213;
\pr {D 46} (1992) 3146
\bibitem{bg} C.W.Bernard and M.F.L.Golterman, \pr {D 46} (1992) 853;
\np {B (Proc.Suppl.) 26} (1992) 360
\bibitem{ks} S.Kim and D.K.Sinclair, \pr {D 52} (1995) 2614
\bibitem{nonpertren1} G.Martinelli at al., \np {B 445} (1995) 81
\bibitem{nonpertren2} K.Jansen et al., \pl {B 372} (1996) 275
\bibitem{mnalmunecar} M.Neubert, These Proceedings
\bibitem{mandula} J.E.Mandula and M.C.Ogilvie, \np {B (Proc.Suppl.) 30}
(1993) 477
\bibitem{aglietti} U.Aglietti, \np {B 421} (1994) 421
U.Aglietti and S.Capitani, \np {B 432} (1994) 315
\bibitem{eichtencapri} E.Eichten, G.Hockney and H.B.Thacker,
\np {B (Proc Suppl) 17} (1990) 529
\bibitem{boucaud} P.Boucaud et al., \pl {B 220} (1989) 219 
\bibitem{eichtensmeared} A.Duncan et al., \np {B (Proc.Suppl.) 34} (1994)
444
\bibitem{lm} G.P.Lepage and P.B.Mackenzie, \pr {D 48} (1993) 2250
\bibitem{fb1overm} S.Collins et al., SCRI preprint, FSU-SCRI-95-92
(1995) \\ (hep-lat/9509065)
\bibitem{fnal} A.Duncan et al., \pr {D 51} (1995) 5101
\bibitem{apefbstat} C.R.Allton, \np {B 437} (1995) 641
\bibitem{sii} H.W.Hamber and C.M.Wu, \pl{B 133} (1983) 351 and
\pl{B 136} (1984) 255; W.Wetzel, \pl{B 136} (1984) 407; T.Eguchi and
N.Kawamoto, \np{B 237} (1984) 572 
\bibitem{heatlie} G.Heatlie et al., \np {B 352} (1991) 266
\bibitem{sw} B.Sheihkoleslami and R.Wohlert, \np {B 259} (1985) 572
\bibitem{naik} R.Wohlert, Ph.D. thesis, unpublished;
S.Naik \pl{B 311} (1993) 230
\bibitem{fnalhyperfine} A.X.El-Khadra, \np {B (Proc.Suppl.) 30} (1993)
449
\bibitem{km} A.S.Kronfeld and P.B.Mackenzie, Ann.Rev.Nucl.Part.Sci.
\underline{43} (1993) 793
\bibitem{perfect} P.Hasenfratz and F.Niedermayer, \np {B 414} (1994) 785
\bibitem{quenchedfd} UKQCD Collaboration, R.M.Baxter et al., \pr {D 49}
(1994) 1594
\bibitem{abada} A.Abada et al., \np {B 376} (1992) 172
\bibitem{static} UKQCD Collaboration, A.K.Ewing et al., Southampton
University preprint SHEP-95-20 (1995) (hep-lat/9508030)
\bibitem{cdf} CDF Collaboration, 
URL;http://www-cdf.fnal.gov/physics/new/\\ 
bottom/cdf3352
\bibitem{baryons} UKQCD Collaboration, K.C.Bowler et al., 
University of Southampton preprint SHEP-95-31 (1995) (hep-lat/9601022)
\bibitem{fhh} J.M.Flynn, O.F.Hern\' andez and B.R.Hill, \pr {D 43} (1991)
3709
\bibitem{bp} A.Borelli and C.Pittori, \np {B385} (1992) 502
\bibitem{draper} T.Draper and C.McNeile, University of Kentucky
preprint, UK-95-10 (1995) (hep-lat/9509060)
\bibitem{bs} C.W.Bernard, T.Draper, G.Hockney and A.Soni,
\pr {D 38} (1988) 3540
\bibitem{jlqcd} JLQCD Collaboration, S.Aoki et al., 
KEK preprint KEK-CP-38 (1995) (hep-lat/9510033) 
\bibitem{bocicchio} C.R.Allton et al., \np {B 372} (1992) 403
\bibitem{gms} V.Gim\' enez, G.Martinelli and C.T.Sachrajda, 
Rome Preprint, in preparation
\bibitem{hyperfineexperiment} Particle Data Group (L.Montanet  et al.),
\pr {D 50} (1994) 1173
\bibitem{ukqcdhyperfine} UKQCD collaboration, C.R.Allton et al., 
\pl {B 292} (1992) 408
\bibitem{neubert} M.Neubert, \prep {245} (1994) 259
\bibitem{renormalons}
G. 't Hooft, in: {\it The Whys of Subnuclear Physics},
ed. A.~Zichichi (Plenum Press, New York, 1979), p.~943;
B. Lautrup, \pl {B69} (1977) 109;
G. Parisi, \pl {B76}  (1978) 65 and
\np {B150} (1979) 163;
F. David, \np {B234} (1984) 237 and
{\it ibid.\/} \underline {B263} (1986) 637;
A.H. Mueller, \np {B250} (1985) 327, \pl {B308}
  (1993) 355 and in the proceedings of the Workshop ``QCD: 20 years
  Later'', Aachen, June 1992, eds. P.M.Zerwas and H.A.Kastrup (World
  Scientific, Singapore, 1993), p.~162;
V.I. Zakharov, \np {B385}, 452 (1992);
M. Beneke and V.I. Zakharov, \prl {69} (1992) 2472;
M. Beneke, \pl {B307} (1993) 154 and \np {B405} (1993) 424;
D. Broadhurst, Z.\ Phys.\ \underline{C58} (1993) 339
\bibitem{repar} M.Luke and A.V.Manohar, \pl {B286} (1992) 348
\bibitem{ms1} G.Martinelli and C.T.Sachrajda \pl {B 354} (1995) 423
\bibitem{ms2} G.Martinelli and C.T.Sachrajda, Cern Preprint
CERN-TH/96-117 \\ (1996) (hep-ph/9605336)
\bibitem{bsuv} I.I.Bigi, M.A.Shifman, N.G.Uraltsev and A.I.Vainshtein, 
\pr{D50} (1994) 2234 
\bibitem{bb} M.Beneke and V.M.Braun, \np{B426} (1994) 301 
\bibitem{cgms} M.Crisafulli, V.Gim\' enez, G.Martinelli and 
C.T.Sachrajda, \np {B 457} (1995) 594
\bibitem{sterman1} H.Contopanagos and G.Sterman, \np {B419} (1994) 77
\bibitem{az} R.Akhoury and V.I.Zakharov, Univ. of Michigan Preprint,
UM-TH-95-19 (1995) (hep-ph/9507253)
\end{thebibliography}
\end{document}